\documentclass[journal]{IEEEtran}

\usepackage{xspace,amsmath,amssymb,amsfonts,epsfig,subfigure,syntonly}
\usepackage{cite,bm,color,url,textcomp,empheq,boxedminipage}
\usepackage{algorithmicx,algorithm}
\usepackage{makecell}
\usepackage{empheq}
\usepackage{pifont}
\usepackage{soul}

\usepackage{graphicx,graphics}  
\usepackage{multirow,multicol}
\usepackage{psfrag}    
\usepackage{stfloats}
\usepackage{url}

\newtheorem{lemma}{\underline{Lemma}}[section]

\newtheorem{proposition}{\underline{Proposition}}[section]

\newtheorem{remark}{\underline{Remark}}[section]

\newcommand{\mv}[1]{\mbox{\boldmath{$ #1 $}}}

\long\def\symbolfootnote[#1]#2{\begingroup
\def\thefootnote{\fnsymbol{footnote}}
\footnote[#1]{#2}\endgroup}

\psfull
\begin{document}
\title{Fundamental Rate Limits of UAV-Enabled Multiple Access Channel with Trajectory Optimization}
\author{Peiming~Li~and~Jie~Xu
\thanks{Part of this paper has been presented at the IEEE/CIC International Conference on Communications in China (ICCC), Chengdu, China, Aug. 11--13, 2019 \cite{iccc}. {\textit{(Corresponding author: Jie Xu.)}}}
\thanks{P. Li is with the School of Information Engineering, Guangdong University of Technology, Guangzhou, 510006 China (e-mail: peiminglee@outlook.com).}
\thanks{J. Xu is with the School of Information Engineering, Guangdong University of Technology, Guangzhou, 510006 China, and also with the National Mobile Communications Research Laboratory, Southeast University, Nanjing 210018, China (e-mail: jiexu@gdut.edu.cn).}
}
\maketitle

\begin{abstract}
This paper studies an unmanned aerial vehicle (UAV)-enabled multiple access channel (MAC), in which multiple ground users transmit individual messages to a mobile UAV in the sky. We consider a linear topology scenario, where these users locate in a straight line and the UAV flies at a fixed altitude above the line connecting them. Under this setup, we jointly optimize the one-dimensional (1D) UAV trajectory and wireless resource allocation to reveal the fundamental rate limits of the UAV-enabled MAC, under the users' individual maximum power constraints and the UAV's maximum flight speed constraints. First, we consider the capacity-achieving non-orthogonal multiple access (NOMA) transmission with successive interference cancellation (SIC) at the UAV receiver. In this case, we characterize the capacity region by maximizing the average sum-rate of all users subject to a set of rate profile constraints. To optimally solve this highly non-convex problem with infinitely many UAV location variables over time, we show that any speed-constrained UAV trajectory is equivalent to the combination of a maximum-speed flying trajectory and a speed-free trajectory, and accordingly transform the original speed-constrained trajectory optimization problem into a speed-free problem that is optimally solvable via the Lagrange dual decomposition. It is rigorously proved that the optimal 1D trajectory solution follows the successive hover-and-fly (SHF) structure, i.e., the UAV successively hovers above a number of optimized locations, and flies unidirectionally among them at the maximum speed. Next, we consider two orthogonal multiple access (OMA) transmission schemes, i.e., frequency-division multiple access (FDMA) and time-division multiple access (TDMA). We maximize the achievable rate regions in the two cases by jointly optimizing the 1D trajectory design and wireless resource (frequency/time) allocation. It is shown that the optimal trajectory solutions still follow the SHF structure but with different hovering locations for each scheme. Finally, numerical results show that the proposed optimal trajectory designs achieve considerable rate gains over other benchmark schemes, and the capacity region achieved by NOMA significantly outperforms the rate regions by FDMA and TDMA.
\end{abstract}
\vspace{2pt}
\section{Introduction}
\IEEEPARstart{U}{nmanned} aerial vehicle (UAV)-enabled communication platforms have emerged as a promising technology in next generation wireless networks, which not only provide basic wireless coverage for remote areas without adequate ground infrastructures, but also enhance the communication rates in temporary hotspots\cite{ZYwireless}. In the industry, various companies have carried out their UAV-enabled wireless communication projects, some examples including Google's Project Loon \cite{g}, Facebook's Project Aquila \cite{fb}, Nokia's Flying-cell (F-Cell)\cite{Nokia}, China Mobile's UAV base stations (BSs) \cite{china}, and Alibaba's Cloud IoT in the Sky \cite{ali}. In the academia, growing research efforts have been devoted to employing UAVs as aerial wireless platforms such as mobile relays \cite{zengT,radio1,Rfan}, BSs \cite{MMW,AA,MA,CZ,MM,PJ,Propla}, wireless chargers \cite{JIEXU,LX}, and mobile edge computing (MEC) servers \cite{12,FZhou}. Different from conventional terrestrial wireless communication infrastructures, UAVs in the sky generally have various advantages. For example, UAVs can be deployed rapidly in a cost-effective manner in emergency situations (e.g., after earthquake disasters) \cite{ZYwireless}. Also, UAVs in the sky possess strong line-of-sight (LoS) wireless links with ground users, which help to provide more reliable communications between them \cite{newzeng}. Furthermore, UAVs can exploit the fully-controllable mobility via positioning adjustment (e.g., \cite{radio1,PJ,AA,Rfan,MA,MM,CZ,Propla,yuwei,bailin}) or trajectory design (e.g., \cite{9,10,11,12,13,zengT,Czhang,JIEXU,LX,FZhou,jcin2}) for optimizing the communication performance.

Among others, the joint optimization of UAV trajectory and wireless resource allocations has received particular research interests to improve the performance of UAV-enabled wireless communications. Generally speaking, by exploiting the mobility, the UAV can fly closer to its communicating ground node to reduce the path loss for efficient communication. Nevertheless, when there are multiple communicating nodes on the ground, how to design the UAV trajectory to balance their rate performance tradeoffs becomes a non-trivial problem. In the literature, there have been several existing works investigating this problem under different setups (see, e.g., \cite{9,10,11,12,13,zengT,Czhang,JIEXU,LX,FZhou,jcin2}). For instance, the authors in \cite{zengT} considered the trajectory optimization for throughput maximization in a UAV-enabled relaying system aided by data buffering, in which the UAV-enabled relay flies between the ground source and destination nodes to efficiently decode, store, and forward messages. Furthermore, \cite{9,10,13,Czhang} studied UAV-enabled multiuser communication systems, where one single UAV acts as a BS or access point (AP) to communicate with multiple ground users via orthogonal frequency-division multiple access (OFDMA) \cite{10,13} and time-division multiple access (TDMA) \cite{9,Czhang}, respectively. By proper trajectory design, the UAV can sequentially visit these users to shorten the transmission distances, thus maximizing different users' communication performance in a fair manner. These results were then extended to multi-UAV-enabled wireless networks in \cite{11}. In this case, multiple UAVs jointly design their trajectories to not only shorten the transmission distances with intended users for better link quality, but also enlarge the distances with undesirable users for interference mitigation.

Despite recent research progress, prior works usually considered low-complexity and suboptimal transmission schemes (e.g., OFDMA or TDMA for multiuser communication \cite{10,13,9,Czhang}), and employed generally suboptimal trajectory optimization approaches (like travelling salesman problem (TSP) and successive convex programming (SCP) \cite{zengT,10,13,9,Czhang,11,FZhou}). There are rare works characterizing the fundamental rate limits of UAV-enabled wireless communications with capacity-achieving transmission strategies and globally optimal trajectory designs. This, however, is very important for understanding the system performance upper bound and guiding practical system designs. In particular, consider the UAV-enabled multiple access channel (MAC) and broadcast channel (BC), in which one UAV communicates with multiple ground users in the uplink and downlink, respectively. By employing capacity-achieving non-orthogonal multiple access (NOMA) transmission strategies (i.e., with successive interference cancellation (SIC) for MAC and with superposition coding together with SIC for BC \cite{dual,david,R10}) at the UAV receiver, how to jointly optimize the UAV trajectory and wireless resource allocation for maximizing the capacity region is a challenging task that has not been well understood.{\footnote{Please refer to \cite{sic2} for the practical implementation of NOMA in UAV communications. However, how to maximize the capacity region of UAV-enabled communications (with capacity-achieving NOMA) has not been addressed in \cite{sic2}.}} To our best knowledge, there is only one prior work \cite{qq} that investigated the capacity region of a UAV-enabled two-user BC (and also MAC due to the uplink-downlink duality \cite{dual}). It was shown that to maximize the capacity region, the optimal UAV trajectory should follow a hover-fly-hover (HFH) structure, i.e., the UAV successively hovers at a pair of initial and final locations above the line segment of the two users with optimized durations and flies unidirectionally between them at the maximum speed, during which superposition coding is generally needed \cite{qq}. It is worth noting that the optimality of the HFH trajectory is verified via a very complicated proof technique based on the ``monotonicity'' of the two-user BC/MAC capacity region under any given UAV locations, which, however, cannot be applied to the general case with more than two ground users. Therefore, how to optimize the UAV trajectory, reveal its optimal structure, and accordingly characterize the capacity region for general UAV-enabled MAC and BC is a difficult problem that has not been addressed yet.

In this paper, we study a UAV-enabled MAC, in which multiple users on the ground transmit individual messages to a UAV flying in the sky. For the purpose of revealing the most essential engineering insights on UAV trajectory design, we consider a special linear topology scenario, where these users locate in a straight line and the UAV flies at a fixed altitude above the line connecting them. In practice, this may correspond to scenarios with linearly deployed users in, e.g., bridges, roads, and tunnels. Under this setup, we jointly optimize the one-dimensional (1D) UAV trajectory and wireless resource allocation to reveal the fundamental rate/capacity regions. Here, the rate/capacity region is defined as the set of average rate tuples over a particular communication period, which are simultaneously achievable by all users, under the users' individual maximum power constraints and the UAV's maximum flight speed constraints. Our results are summarized as follows.
\begin{itemize}
\item First, we consider the capacity-achieving NOMA transmission with SIC at the UAV receiver, based on which we characterize the capacity region by maximizing the average sum-rate of all users subject to a set of rate profile constraints. The sum-rate maximization problem, however, is highly non-convex and consists of infinitely many UAV location variables over continuous time, which is thus very difficult to be optimally solved via conventional approaches. Despite this difficulty, we present the globally optimal solution to this problem, by showing that any speed-constrained UAV trajectory is equivalent to the combination of a maximum-speed flying trajectory and a speed-free trajectory, and accordingly transforming the original speed-constrained trajectory optimization problem into a speed-free problem that is optimally solvable via the Lagrange dual decomposition. It is rigorously proved that the optimal 1D trajectory solution follows the successive hover-and-fly (SHF) structure, i.e., the UAV successively hovers above a number of optimized locations, and flies unidirectionally among them at the maximum speed. During the flight, the UAV needs to properly design the decoding orders for these users, for which time-sharing is generally required.
\item Next, we consider two orthogonal multiple access (OMA) transmission, i.e., frequency-division multiple access (FDMA) and TDMA. For both cases, we optimize the UAV trajectory, jointly with the wireless resource (bandwidth/time) allocations over time, to maximize the achievable rate regions based on the rate profile technique. Similarly as in the NOMA case, we obtain the globally optimal solutions to the two rate maximization problems by first transforming the original problems into speed-free trajectory optimization problems and then using the Lagrange dual decomposition. It is shown that the optimal UAV trajectory solutions still follow the SHF structure but with different hovering locations for each scheme.
\item Finally, we present numerical results to validate the performance of our proposed designs. It is shown that the optimal trajectory design achieves considerable rate gains over other benchmark schemes. It is also shown that the capacity region achieved by NOMA significantly outperforms the rate regions by OMA, while FDMA achieves higher rate region than that by TDMA. When the flight duration becomes large, it is shown that for NOMA and FDMA, the UAV's hovering locations are generally above middle points among ground users; while for TDMA, the UAV's hovering locations should be exactly above the corresponding communicating ground users.
\end{itemize}

The remainder of this paper is organized as follows. Section \ref{sec:II} introduces the system model of the UAV-enabled MAC and formulates the capacity region characterization problem under NOMA. Section \ref{sec:III} presents the optimal solution to the capacity region characterization problem. Section \ref{V+} studies the rate region maximization under FDMA and TDMA. Section \ref{sec:V} provides numerical results to demonstrate the performance of our proposed designs versus benchmark schemes. Finally, Section \ref{sec:VI} concludes this paper and discusses future work.
\begin{figure}[t]
\centering
    \includegraphics[width=8cm]{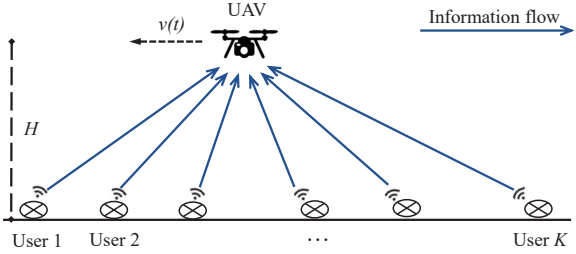}
\caption{Illustration of the UAV-enabled MAC.} \label{f1}
\end{figure}
\section{System Model}\label{sec:II}
In this paper, we consider a UAV-enabled MAC, in which $K > 1$ ground users send individual messages to a mobile rotary-wing UAV in the sky, over a finite communication period ${\cal T}\triangleq (0,T]$ with duration $T>0$,{\footnote{In practice, the UAV mission/communication duration $T$ cannot exceed the UAV's maximum battery lifetime. For example, the maximum flight duration of the DJI's Mavic 2 UAV is approximately 31 minutes (see {\url{https://www.dji.com/cn/mavic-2-enterprise/info#specs}}), and that of the DJI's T16 UAV is approximately 18 minutes (see {\url{ https://www.dji.com/cn/t16/info#specs}}).}} as shown in {Fig.~\ref{f1}}. We focus on the linear topology scenario, where all the ground users are deployed in a straight line with altitude zero. Let $(w_k,0)$ denote the location of each ground user $k\in{\cal K}\triangleq \{1,\ldots,K\}$ in a two-dimensional (2D) Cartesian coordinate system. Here, we assume $w_1 \le \ldots \le w_K$ without loss of generality. It is also assumed that the UAV flies above the line connecting these users at a fixed altitude $H > 0$, with the time-varying location being $\left(x(t),H\right)$ at time instant  $t\in\mathcal T$. The distance between the UAV and
user $k$ at time instant  $t\in\mathcal T$ is given by $d_k(t)=\sqrt{(x(t)-w_k)^2+H^2}$, $k\in{\cal K}$. Let $V_{\textrm{max}}\geq 0$ denote the maximum UAV flight speed in meters/second (m/s). We thus have
\begin{align}\label{1}
|{\dot{x}}(t)|\leq V_{\textrm{max}},\ \forall t\in{\cal T},
\end{align}
where ${\dot{x}}(t)$ denotes the first-order derivative of $x(t)$ with respect to time $t$. For the ease of reading, the notations used in this paper are listed in Table \ref{t1}.
\begin{table}[ht]
\caption{List of Notations and the Corresponding Physical Meanings}\label{t1}
\centering
\begin{tabular}{|l|p{6.77cm}|}
\hline
$\cal K$&Set of ground users\\ \hline
$K$&Number of ground users\\ \hline
$\cal T$&Communication period\\ \hline
$T$&Communication duration\\ \hline
${\bar T}$&Duration for the maximum-speed flying trajectory\\ \hline
${\hat T}$&Duration for the speed-free trajectory\\ \hline
$\{x(t)\}$&UAV's speed-constrained trajectory\\ \hline
$\{\bar{x}(t)\}$&UAV's maximum-speed flying trajectory\\ \hline
$\{\hat{x}(t)\}$&UAV's speed-free trajectory\\ \hline
$w_k$&Horizontal location of ground user $k$\\ \hline
$x_\textrm{I}/x_\textrm{F}$&Initial/final location of the UAV\\ \hline
$V_{\textrm{max}}$&UAV's maximum flight speed\\ \hline
$H$&UAV's flight altitude\\ \hline
$\sigma^2$&Noise power\\ \hline
$d_k(t)$&Distance between the UAV and user $k$ at time $t$\\ \hline
$h_k(t)$&Channel power gain between the UAV and user $k$ at time $t$\\ \hline
$p_k(t)$&Transmit power of user $k$ at time $t$\\ \hline
$r_k$&Average achievable rate of user $k$\\ \hline
${\mv \alpha}$&Rate-profile vector\\ \hline
$\bar{{\mathcal C}}$&Capacity region achieved by NOMA\\ \hline
$\mv \pi$&Decoding order permutation at the UAV for NOMA\\ \hline
$b_k(t)$&Bandwidth allocation for user $k$ at time $t$ under FDMA\\ \hline
${\bar {\cal R}}_{\textrm{FDMA}}$&Achievable rate region by FDMA\\ \hline
$\rho_k(t)$&Time allocation for user $k$ at time $t$ under TDMA\\ \hline
${\bar {\cal R}}_{\textrm{TDMA}}$&Achievable rate region by TDMA\\ \hline
\end{tabular}
\end{table}

Suppose that the transmission from the ground users to the UAV is operated over bandwidth $B$ in Hertz (Hz), where  $T_s=1/B$ denotes the corresponding symbol duration in second (s). We consider that the UAV's location change within each symbol duration is negligible as compared to the UAV's flight altitude $H$, i.e., $V_{\textrm{max}}T_s\ll H$ \cite{qq}. Therefore, the wireless channel from each user to the UAV is invariable within each symbol interval. We consider the probabilistic LoS model for ground-to-air wireless channels, where the LoS probability depends on the elevation angle (see, e.g., \cite{AA,Propla}). In this case, we focus on the average path loss and ignore the shadowing and small-scale fading for the purpose of exposition. Accordingly, the average channel power gain from each ground user $k\in\mathcal K$ to the UAV at time instant $t\in\mathcal T$ is modeled as \cite{AA}
\begin{align}\label{3}
h_k(t)=P_{k,\textrm{LoS}}(t)\beta_0 d_k^{-{\epsilon}}(t)+\xi(1-P_{k,\textrm{LoS}}(t))\beta_0 d_k^{-\epsilon}(t),
\end{align}
where $P_{k,\textrm{LoS}}(t)$ denotes the LoS probability, $\beta_0$ denotes the path loss at the reference distance of $d_0=1$ m, $\epsilon$ denotes the path loss exponent, $\xi<1$ denotes the additional attenuation factor due to the non-LoS condition. In particular, $P_{k,\textrm{LoS}}(t)$ can be modeled as a logistic function with respect to the elevation angle $\theta_k(t)=({180}/{\pi})\sin^{-1}\left({H}/{d_k(t)}\right)$ in degree \cite{AA}, which is given as $P_{k,\textrm{LoS}}(t)={(1+C\exp(-D(\theta_k(t)-C)))}^{-1}$,
where $C$ and $D$ are parameters determined by the propagation environment. Notice that in the special case with $P_{k,\textrm{LoS}}(t)= 1$ and $\epsilon = 2$, the average path loss model in \eqref{3} corresponds to the simplified free-space path loss model that has been widely adopted in UAV trajectory designs (see, e.g., \cite{qq,zengT,9,13,11}).

At time instant $t\in{\cal T}$, let $s_k(t)$ denote the information-bearing signal transmitted by user $k\in\mathcal K$. Accordingly, the received signal at the UAV is expressed as
\begin{align}\label{2}
y(t)=\sum\limits_{k\in\mathcal K }\sqrt{h_k(x(t))}s_k(t)+n(t),
\end{align}
where $n(t)$ denotes the additive white Gaussian noise (AWGN) at the UAV receiver with mean zero and variance $\sigma^2$. Under given UAV trajectory $\{x(t)\}$, the signal model in \eqref{2} corresponds to a conventional fading MAC with $K$ transmitters (ground users) communicating with one receiver (UAV) \cite{david}. In order to achieve the capacity region of this channel, the ground users should employ Gaussian signaling by setting $s_k(t)$'s as independent circularly symmetric complex Gaussian (CSCG) random variables with mean zero and variances $\mathbb{E}(|s_k(t)|^2) = p_k(t), \forall k\in{\cal K}$, where $\mathbb{E}(\cdot)$ denotes the statistical expectation, and $p_k(t)$ denotes user $k$'s transmit power at time $t$. Suppose that at each time instant $t$, each user $k$ is subject to a maximum power constraint $P$,{\footnote{In order to focus on the trajectory optimization for capacity characterization of the UAV-enabled MAC, we consider the instantaneous power constraints at each user, similarly as in \cite{qq}. Our design principles can be extended to the case with average power constraints at ground users, for which adaptive power allocation over time should be considered.}} i.e.,
\begin{align}
0\le p_k(t) \le P,\ \forall  k\in {\cal K},t\in {\cal T}.
\end{align}
To achieve the channel capacity, the UAV adopts SIC to decode the messages from the $K$ users. Let the permutation $\mv \pi = [\pi(1),\ldots,\pi(K)]$ denote the decoding order at the UAV, which indicates that the UAV receiver first decodes user $\pi(K)$'s message $s_{\pi(K)}(t)$, then decodes user $\pi(K-1)$'s message $s_{\pi(K-1)}(t)$ by canceling the interference from $s_{\pi(K)}(t)$, followed by $s_{\pi(K-2)}(t)$, $s_{\pi(K-3)}(t)$, and so on, until $s_{\pi(1)}(t)$. By considering the maximum power transmission for rate maximization with $p_k(t) = P, \forall k\in {\cal K},t\in {\cal T}$, the achievable rate in bits per second per Hertz (bps/Hz) at user $\pi(k), k\in {\cal K}$, is given by
\begin{align}\label{44}
r_{\pi(k)}=\frac{1}{T}\int_0^T\log_2\left(\frac{\sigma^2+
\sum_{i=1}^{k}Ph_{\pi(i)}(x(t))}{\sigma^2+\sum_{i=1}^{k-1}Ph_{\pi(i)}(x(t))}\right)\textrm{d}t.
\end{align}

By properly designing the decoding order and allowing time-sharing among different decoding orders, the region of all achievable average rate tuples $\mv r = [r_1,\ldots, r_K]$ in bps/Hz for the $K$ ground users under given $\{x(t)\}$ is expressed as \cite{david}
\begin{align}
 \label{4}\bar{{\mathcal C}}&(\left\{x(t)\right\})=\left\{\mv r\in{\mathbb R}^{+}_K \bigg|
 \sum_{k\in \bar{\mathcal K}}r_k \leq\right.\notag\\ &~~~\left.\frac{1}{T}\!\int^T_0\!\log_2\left(1+\sum_{k\in \bar{\mathcal K}}\frac{Ph_k(x(t))}{\sigma^2}\right)\textrm{d}t,\ \forall\bar{\mathcal K}\subseteq{\cal K}\right\},
\end{align}
where ${\mathbb R}^{+}_K$ denotes the set of all non-negative real vectors with dimension $K$.

Let $\mathcal{X}$ denote the feasible set of $\left\{x(t)\right\}$ specified by the UAV's maximum speed constraints in \eqref{1}. Then the capacity region of the UAV-enabled MAC is defined as
\begin{align}
{\mathcal C}(V_{\textrm{max}},T)=~~~\bigcup_{\mathclap{\left\{x(t)\right\}\in{\mathcal X}}}~~~\bar{{\mathcal C}}(\left\{x(t)\right\}),
\end{align}
which consists of all the achievable average rate tuples for the ground users over the communication period $\cal T$, subject to the UAV's maximum speed constraint in \eqref{1}.

In this paper, we are interested in characterizing the Pareto (or the upper-right) boundary of the capacity region ${\mathcal C}(V_{\textrm{max}}, T)$, at which each user cannot increase its achievable average rate unless sacrificing the rates of other users. Specifically, let ${\mv \alpha}=[\alpha_1,\ldots,\alpha_K]$ denote a rate-profile vector that specifies the rate allocation among the $K$ ground users with $\alpha_k\geq 0, \forall k\in{\cal K}$, and $\sum_{k\in{\cal K}}\alpha_k=1$. Here, a larger value of $\alpha _k$ means that ground user $k$ has a higher communication priority to achieve a larger data rate. Then, the characterization of any Pareto boundary point of the capacity region is formulated as the following optimization problem:
\begin{align}
\textrm{(P1):}~&\max_{\left\{x(t)\right\},{\mv r},{R}}
~{R}\notag\\
&~~~~{\textrm{s.t.}}
~r_k\geq\alpha_{k}{R},\ \forall k\in{\cal K}\label{5}\\
&~~~~~~~~~{\mv r}\in\bar{{\cal C}}(\left\{x(t)\right\})\label{6}\\
&~~~~~~~~~|\dot{x}(t)|\leq V_{\textrm{max}},\ \forall t\in \cal T,\label{7}
\end{align}
where $R$ denotes the average achievable sum-rate of the $K$ ground users. Problem (P1) is generally a highly non-convex optimization problem that contains infinitely many variables over continuous time.
\vspace{1pt}
\begin{remark}
Notice that when there are only two users with $K=2$, problem (P1) is simplified as
\begin{align}
&\max_{\left\{x(t)\right\},r_1,r_2,{R}}
~{R}\label{P2}\\
&~~~~~~{\textrm{s.t.}}~
r_k\geq\alpha_{k}{R},\ \forall k\in\{1,2\}\notag\\
&~~~~~~~~~~~r_k\le\log_2\left(1+\frac{Ph_k(x(t))}{\sigma^2}\right),\ \forall k\in\{1,2\}\notag\notag\\
&~~~~~~~~~~~r_1+r_2\le\log_2\left(1+\frac{P(h_1(x(t))+h_2(x(t)))}{\sigma^2}\right)\notag\\
&~~~~~~~~~~~|\dot{x}(t)|\leq V_{\textrm{max}},\ \forall t\in {\cal T}.\notag
\end{align}
It is evident that due to the uplink-downlink duality, problem \eqref{P2} corresponds to the capacity region characterization problem for the UAV-enabled two-user BC/MAC in \cite{qq} with fixed power allocation. It thus follows directly from \cite{qq} that the optimal UAV trajectory solution to problem \eqref{P2} has the so-called HFH structure, i.e., the UAV first hovers at the initial location $x_\textrm{I}$ for duration $t_\textrm{I}$, then flies unidirectionally to the final location $x_\textrm{F} \ge x_{\textrm{I}}$ at the maximum speed $V_\textrm{max}$, and finally hovers at $x_\textrm{F}$ for the remaining duration $t_\textrm{F} = T - t_\textrm{I}-(x_\textrm{F}-x_\textrm{I})/V_\textrm{max}$. However, the optimality of the HFH trajectory is proved in \cite{qq} based on the ``monotonicity'' of the capacity region under any given UAV locations, which cannot be extended to solve problem (P1) in the general UAV-enabled MAC with $K > 2$ ground users. Therefore, problem (P1) is much more difficult to be optimally solved than problem \eqref{P2}.
\end{remark}
\section{Optimal Solution to Problem (P1)}\label{sec:III}
In this section, we propose an efficient approach to $\textrm{optimally}$ solve problem (P1). Before proceeding, we first present the following lemma, which follows directly from \cite[lemma 3]{qq}.
\vspace{1pt}
\begin{lemma}\label{le1}
There always exists a unidirectional UAV trajectory that is optimal for problem (P1), i.e.,
\begin{align*}
x(t_1)\le x(t_2),\ \forall t_1, t_2 \in {\cal T}, t_1 \le t_2.
\end{align*}
\end{lemma}

Based on Lemma \ref{le1}, we focus on the unidirectional trajectory to problem (P1) without loss of optimality. Suppose that the initial and final locations of the trajectory are denoted by $x(0) = x_\textrm{I}$ and $x(T) = x_\textrm{F}$, respectively, which are optimization variables to be decided later. Here, it must follow that $w_1\le x_{\textrm I} \le x_{\textrm F}\le w_K$, such that the UAV always flies within the line segment above the $K$ ground users' locations to maximize the capacity region. In the following, we solve problem (P1) by first solving the following problem (P1.1) under any given initial and final locations $x_{\textrm I}$ and $x_{\textrm F}$, and then using a 2D exhaustive search over $[w_1, w_K] \times [w_1,w_K]$ to find the optimal initial and final locations $x_{\textrm I}$ and $x_{\textrm F}$. Notice that under arbitrarily chosen initial and final locations $x_\textrm{I}$ and $x_\textrm{F}$, problem (P1.1) is generally not equivalent to problem (P1); however, under the optimal initial and final locations $x_\textrm{I}$ and $x_\textrm{F}$ obtained via the 2D exhaustive search, problems (P1.1) and (P1) become equivalent.
\begin{align}
\textrm{(P1.1):}~&\max_{\left\{x(t)\right\},{\mv r},{R}}
~{R}\notag\\
&~~~~{\textrm{s.t.}}~\eqref{5}, ~\eqref{6},~\eqref{7}\notag\\
&~~~~~~~~~x_{\textrm{I}}\le {x}(t)\leq x_{\textrm{F}},\ \forall t\in {\cal T}.\label{8}
\end{align}

In the rest of this section, we will focus on solving problem (P1.1) under any given initial and final locations $x_{\textrm I}$ and $x_{\textrm F}$ with $w_1\le x_{\textrm I} \le x_{\textrm F}\le w_K$. In particular, we first reformulate the trajectory optimization problem (P1.1) with speed constraints in \eqref{7} as an equivalent speed-free trajectory optimization problem, and then employ the Lagrange duality method to obtain the optimal solution.

\subsection{Problem Reformulation of Problem (P1.1)}
First, we make the following definitions for notational convenience.
\begin{itemize}
\item{{\it{Speed-constrained}} trajectory $\{x(t)\}_{t=0}^{T}$:} This corresponds to our original duration-$T$ trajectory with the maximum speed constraint $V_\textrm{max}$ over time, i.e., $|\dot{x}(t)|\le V_{\textrm{max}},\forall t\in {\cal T}$.
\item{{\it{Maximum-speed}} flying trajectory $\{\bar{x}(t)\}_{t=0}^{\bar T}$:} In this trajectory, the UAV flies from the initial location $x_\textrm{I}$ to the final location $x_\textrm{F}$ at the maximum speed $V_{\textrm{max}}$, with duration $\bar{T} = ({x_{\textrm{F}}\!-\!x_{\textrm{I}}})/V_{\textrm{max}}$. Notice that under any fixed $x_\textrm{I}$ and $x_\textrm{F}$, the maximum-speed flying trajectory $\{\bar x(t)\}_{t=0}^{\bar T}$ is fixed, given by
    \begin{align}
    \bar{x}(t)=x_{\textrm{I}}+V_{\textrm{max}}t,\ \forall t\in(0,\bar{T}].\label{9}
    \end{align}
\item{{\it{Speed-free}} trajectory $\{\hat{x}(t)\}_{t=0}^{\hat T}$:} In this trajectory with duration $\hat{T} = T - {\bar T}$, the UAV can arbitrarily adjust its location over time without any speed constraints. For instance, the UAV can hover at different locations at two consecutive time instants,  without spending any time for flying between them. Notice that the speed-free trajectory is only a mathematic equivalence that is used for helping solve problem (P1.1), but generally not implementable in practice.
\end{itemize}

Then, we have the following lemma.
\vspace{1pt}
\begin{lemma}\label{l2}
For any given speed-constrained trajectory $\{x(t)\}_{t=0}^T$ with initial and final locations $x_\textrm{I}$ and $x_\textrm{F}$, and $T \ge (x_\textrm{F}-x_\textrm{I})/V_\textrm{max}$, we can always construct a maximum-speed flying trajectory $\{\bar{x}(t)\}_{t=0}^{\bar T}$ with duration $\bar{T}=(x_\textrm{F}-x_\textrm{I})/V_\textrm{max}$ and a speed-free trajectory $\{\hat{x}(t)\}_{t=0}^{\hat T}$ with duration $\hat{T}=T-\bar{T}$, such that the combination of $\{\bar{x}(t)\}$ and $\{\hat{x}(t)\}$ achieves the same capacity region that is achievable by $\{x(t)\}$. In other words, let ${\hat {\mathcal C}}(\{\bar x(t)\},\{\hat x(t)\})$ denote the rate region achieved by the combination of $\{\bar{x}(t)\}$ and $\{\hat{x}(t)\}$, given by \eqref{111} at the top of next page.
\begin{figure*}[!t]
\begin{align} \label{111}
&{\hat {\mathcal C}}(\{\bar x(t)\},\{\hat x(t)\}) =\notag\\ &~~~~\left\{\mv r\in\mathbb{R}^+_K \bigg| \sum_{k\in \bar{\mathcal K}}r_k\leq\frac{1}{T}\left(\int^{\bar{T}}_0\!\log_2\left(1+\sum\limits_{k\in \bar{\mathcal K}}\frac{Ph_k(\bar{x}(t))}{\sigma^2}\right)\textrm{d}t+\int^{\hat{T}}_0\!\log_2\left(1+\sum\limits_{k\in \bar{\mathcal K}}\frac{Ph_k(\hat{x}(t))}{\sigma^2}\right)\textrm{d}t\right),\ \forall\bar{\mathcal K}\subseteq{\cal K}\right\}.
\end{align}
\end{figure*}
Then we have
\begin{align}
{\bar {\mathcal C}}(\{x(t)\}) = {\hat {\mathcal C}}(\{\bar x(t)\},\{\hat x(t)\}).
\end{align}
\end{lemma}
\begin{IEEEproof}
See Appendix \ref{AppA}.
\end{IEEEproof}

Based on Lemma \ref{l2}, it is evident that the optimization of speed-constrained trajectory $\{x(t)\}$ in problem (P1.1) is equivalent to optimizing a maximum-speed flying trajectory $\{\bar{x}(t)\}$ and a speed-free trajectory $\{\hat{x}(t)\}$. Notice that as $x_{\textrm{I}}$ and $x_{\textrm{F}}$ are fixed, the maximum-speed flying trajectory $\{\bar{x}(t)\}$ is actually fixed, as given in \eqref{9}. Therefore, we only need to optimize the speed-free trajectory $\{\hat{x}(t)\}$. By replacing $\bar{\mathcal C}(\{x(t)\})$ as ${\hat {\mathcal C}}(\{\bar x(t)\},\{\hat x(t)\})$, the speed-constrained trajectory optimization problem (P1.1) is equivalently reformulated as the following speed-free trajectory optimization problem:
\begin{align}
\textrm{(P1.2):}~&\max_{\left\{\hat{x}(t)\right\},{\mv r},{R}}
~{R}\notag\\
&~~~~{\textrm{s.t.}}~\eqref{5}\notag\\
&~~~~~~~~~{\mv r}\in\hat{{\cal C}}\left(\left\{\bar{x}(t)\right\},\{\hat{x}(t)\}\right)\label{17}\\
&~~~~~~~~~x_{\textrm{I}}\le\hat{x}(t)\le x_{\textrm{F}},\ \forall t\in (0,\hat T].\label{18}
\end{align}
Let $\{\hat{x}^\star(t)\}$, $\mv r^\star = [r_1^\star,\ldots, r_K^\star]$, and $R^\star$ denote the optimal solution to problem (P1.2). Accordingly, $\mv r^\star$ and $R^\star$ are also the optimal solution to problem (P1.1), and by combining $\{\hat{x}^\star(t)\}$ together with the maximum-speed flying trajectory $\{\bar{x}(t)\}$ in \eqref{9}, we can also construct the optimal speed-constrained trajectory $\{{x}^\star(t)\}$ to problem (P1.1) as explained in Appendix A. Therefore, we only need to focus on solving problem (P1.2).

\subsection{Optimal Solution to Reformulated Problem (P1.2)}
Although problem (P1.2) is still non-convex, it can be shown to satisfy the so-called time-sharing condition \cite{time}. Therefore, the strong duality holds between problem (P1.2) and its Lagrange dual problem. As a result, we can optimally solve problem (P1.2) by applying the Lagrange duality method \cite{boyd}.

Let $\lambda_k \ge 0$ denote the dual variable associated with the $k$-th rate-profile constraint in \eqref{5}, $k\in {\cal K}$. Then the partial Lagrangian of problem (P1.2) is
\begin{align}
{\cal L}_1(\{\lambda_k\},\{\hat{x}(t)\},{\mv r},R)=(1-\sum\limits_{k\in{\cal K}}\lambda_k \alpha_k){R}+\sum\limits_{k\in{\cal K}}\lambda_k r_k.
\end{align}
Accordingly, the Lagrange dual function of problem (P1.2) is
\begin{align}
f_1({\{\lambda_k\}})=&\max_{\mv r,\{\hat{x}(t)\},R}~{\cal L}_1(\{\lambda_k\},\{\hat{x}(t)\},\mv r,R)\label{15.1}\\
&~~~~{\textrm{s.t.}}~\eqref{17},~\eqref{18}.\notag
\end{align}
\begin{lemma}\label{l3}
In order for the dual function $f_1(\{\lambda_k\})$ to be upper-bounded from above (i.e., $f_1(\{\lambda_k\})<\infty)$, it must hold that $\sum\limits_{k\in{\cal K}}\lambda_k \alpha_k=1$.
\end{lemma}
\begin{IEEEproof}
Suppose that $\sum\limits_{k\in{\cal K}}\lambda_k \alpha_k>1$ or $\sum\limits_{k\in{\cal K}}\lambda_k \alpha_k<1$. Then by setting $R\rightarrow-\infty$ or $R\rightarrow\infty$, we have $f_1(\{\lambda_k\})\rightarrow\infty$. Therefore, this lemma is proved.
\end{IEEEproof}
According to Lemma \ref{l3}, the dual problem of problem (P1.2) is given by
\begin{align}
\textrm{(D1.2):}~&\min_{\{\lambda_k \ge 0\}}~f_1(\{\lambda_k\})\notag\\
&~~{\textrm{s.t.}}
~\sum\limits_{k\in{\cal K}}\lambda_k \alpha_k=1\label{21}.
\end{align}
As the strong duality holds, we can solve problem (P1.2) by equivalently solving its dual problem (D1.2). In the following, we first evaluate $f_1(\{\lambda_k\})$ in \eqref{15.1} under any given $\{\lambda_k\}$, and then solve problem (D1.2) to find the optimal ${\{\lambda_k\}}$, denoted by $\{\lambda_k^\star\}$.

\subsubsection{Evaluating $f_1(\{\lambda_k\})$ by Solving Problem (\ref{15.1})}
First, we obtain the dual function $f_1(\{\lambda_k\})$ under given ${\{\lambda_k\}}$ by solving problem \eqref{15.1}. As $\sum_{k\in {\cal K}}\lambda_k\alpha_k = 1$, the optimal solution of $R^*$ to problem \eqref{15.1} can be chosen as any real value. We have $R^* = 0$ here for obtaining $f_1(\{\lambda_k\})$ only. Therefore,  problem \eqref{15.1} is reduced as
\begin{align}
&\max_{\{\hat{x}(t)\},\mv r}~\sum\limits_{k\in{\cal K}}\lambda_k r_k\label{19}\\
&~~\textrm{s.t.}~\eqref{17},~\eqref{18}.\notag
\end{align}
To solve problem \eqref{19}, we have the following lemma from \cite{david}.
\vspace{1pt}
\begin{lemma}\label{l4}
For any given $\{\lambda_k\}$, the optimal solution to problem \eqref{19} is obtained by a \textit{vertex} $\hat{\mv r}_{\pi}\triangleq\left[\hat r_{\pi(1)},\ldots,\hat r_{\pi(K)}\right]$ of the polymatroid $\hat{{\cal C}}(\left\{\hat{x}(t)\right\},\{\bar{x}(t)\})$, where $\hat r_{\pi(k)}$ is given as
\begin{align}
\hat r_{\pi(k)}=&\frac{1}{T}\int_0^{\bar T}\log_2\left(\frac{\sigma^2+
\sum_{i=1}^{k}Ph_{\pi(i)}({\bar x}(t))}{\sigma^2+\sum_{i=1}^{k-1}Ph_{\pi(i)}({\bar x}(t))}\right)\textrm{d}t\ +\notag\\
&\frac{1}{T}\int_0^{\hat T}\log_2\left(\frac{\sigma^2+
\sum_{i=1}^{k}Ph_{\pi(i)}({\hat x}(t))}{\sigma^2+\sum_{i=1}^{k-1}Ph_{\pi(i)}({\hat x}(t))}\right)\textrm{d}t,\label{eqn:new:XJ1}
\end{align}
where the permutation ${\mv \pi}=\left[\pi(1),\ldots,\pi(K)\right]$ corresponds to the decoding order that is determined such that $\lambda_{\pi(1)}\ge\ldots\ge\lambda_{\pi(K)}\ge0$.
\end{lemma}

Based on Lemma \ref{l4} and substituting \eqref{eqn:new:XJ1}, problem \eqref{19} or problem \eqref{15.1} is reformulated as problem \eqref{25.1} at the top of next page,
\begin{figure*}[!t]
\begin{align} \label{25.1}
\max_{\{x_\textrm{I} \le \hat{x}(t)\le x_\textrm{F}\}} ~\sum\limits_{k\in {\cal K}}\frac{\lambda_{\pi(k)}-\lambda_{\pi(k+1)}}{{T}}\left(\int_{0}^{\hat{T}}\log_2\left(1+\sum\limits_{i=1}^{k}\frac{Ph_{\pi(i)}(\hat{x}(t))}{\sigma^2}\right)\textrm{d}t+ \int_{0}^{\bar{T}}\log_2\left(1+\sum_{i=1}^{k}\frac{Ph_{\pi(i)}(\bar{x}(t))}{\sigma^2}\right)\textrm{d}t\right).
\end{align}
\end{figure*}
where $\lambda_{\pi(K+1)}\triangleq0$ is defined for notational convenience. Notice that by dropping the constant terms, problem \eqref{25.1} can be decomposed into a number of subproblems in \eqref{25.2}, each corresponding to optimizing $\hat{x}(t)$ for time instant $t \in (0,\hat{T}]$.
\begin{align}
&\max_{x_\textrm{I} \le \hat{x}(t)\le x_\textrm{F}}~\psi(\hat{x}(t))\triangleq \notag\\&\sum\limits_{k\in {\cal K} }\frac{\lambda_{\pi(k)}\!-\!\lambda_{\pi(k+1)}}{T}\left(\log_2\!\left(1\!+\!\sum\limits_{i=1}^{k}\frac{Ph_{\pi(i)}(\hat{x}(t))}{\sigma^2}\right)\right)\label{25.2}.
\end{align}
It is worth noting that each subproblem in \eqref{25.2} is identical for different time instant $t\in (0,\hat T]$. As a result, we can adopt a 1D exhaustive search over the region $[x_{\textrm{I}},x_{\textrm{F}}]$ to find the optimal $\hat{x}$, denoted by $\hat{x}^{*}$, which maximizes $\psi(\hat{x})$ subject to $x_{\textrm{I}} \le \hat{x}^* \le x_{\textrm{F}}$. Accordingly, the optimal solution to problem \eqref{25.1} is given by
\begin{align}
\hat{x}^{*}(t)=\hat{x}^{*},\ \forall t\in (0,\hat{T}].
\end{align}
Note that the optimal solution of $\hat{x}^{*}$ is generally non-unique, and we can arbitrarily choose any one of them for obtaining the dual function $f_1(\{\lambda_k\})$ only. Accordingly, we can obtain $\mv r^* = [r_1^*,\ldots, r_K^*]$ by using Lemma 3.4 and \eqref{eqn:new:XJ1}. Therefore, the dual function $f_1(\{\lambda_k\})$ is finally obtained.

\subsubsection{Finding Optimal $\{\lambda^\star_k\}$ to Solve Problem (D1.2)} $\textrm{Next}$, with $f_1(\{\lambda_k\})$ obtained, we search over $\{\lambda_k\}$ to minimize $f_1(\{\lambda_k\})$ for solving problem (D1.2). Since the dual function $f_1(\{\lambda_k\})$ is always convex but in general non-differentiable, we can use subgradient-based methods, such as the ellipsoid method \cite{boyd}, to obtain the optimal ${\{\lambda_k^\star\}}$. Note that the subgradient of the objective function $f_1(\{\lambda_k\})$ is ${\mv s_0}(\{\lambda_k\})=\left[r^*_1,\ldots,r^*_K\right]$, while the equality constraint \eqref{21} is equivalent to two inequality constraints $1-\sum\limits_{k\in{\cal K}}\lambda_k \alpha_k\le 0$ and $-1+\sum\limits_{k\in{\cal K}}\lambda_k \alpha_k\le 0$, with subgradients being ${\mv s_1}(\{\lambda_k\})=-{\mv \alpha}$ and ${\mv s_2}(\{\lambda_k\})={\mv \alpha}$, respectively.

\subsubsection{Constructing Optimal Primal Solution to (P1.2)} Under the optimal dual solution ${\{\lambda_k^{\star}\}}$ to problem (D1.2), the corresponding optimal solution of $\{\hat{x}^{*}(t)\}$, $\mv r^*$, and $R^*$ to problem \eqref{15.1} may not be unique in general. Therefore, we need an additional step to reconstruct the optimal primal solution to problem (P1.2). In particular, suppose that under the optimal dual solution $\{\lambda_k^{\star}\}$, problem \eqref{25.2} has $\Gamma\ge 1$ optimal solutions, denoted by $\{\hat{x}_{\gamma}^\star\}^{\Gamma}_{\gamma=1}$, with $\hat{x}^\star_1 \le \cdots \le \hat{x}^\star_{\Gamma}$. In this case, to obtain the optimal primal solution to problem (P1.2), we need to time-share the $\Gamma$ solutions by allowing the UAV to hover at each of the $\Gamma$ locations $\{\hat x_\gamma^\star\}_{\gamma = 1}^\Gamma$ for a certain duration that needs to be optimized.

On the other hand, it is also worth emphasizing that if there exist some $\lambda^{\star }_k$'s that are equal to each other, then the decoding order at the UAV receiver and the corresponding average rate tuples at users are also non-unique, as shown in Lemma \ref{l4}. In this case, the UAV needs to further time-share among different decoding orders \cite{R10}. Let ${\cal J}_1,\ldots,{\cal J}_M\subseteq{\cal K}$ denote $M$ disjoint subsets such that $\lambda_j^{\star}$'s are identical, $j\in{\cal J}_m$, for any $1\le m\le M$ and $|{\cal J}_m|\ge2$. Define set ${\cal I}\triangleq\left\{1,\ldots,\prod_{m=1}^M|{\cal J}_m|\right\}$. As a result, under the optimal dual solution $\{\lambda^{\star}_k\}$, problem \eqref{15.1} admits a number of $|{\cal I}|$ optimal decoding orders, denoted by ${\mv \pi}^{(1)},\ldots,{\mv \pi}^{(|{\cal I}|)}$.

By combining the $\Gamma$ hovering locations and the $|{\cal I}|$ decoding orders, we have $\Gamma\cdot|{\cal I}|$ associated average rate tuples, denoted by ${\mv r}^{(i)}(\hat{x}^\star_\gamma)$'s, $\forall i\in\{1,\ldots,|{\cal I}|\}, \gamma\in\{1,\ldots,\Gamma\} $, where ${\mv r}^{(i)}(\hat{x}^\star_\gamma)=\left[r_1^{(i)}(\hat{x}^\star_\gamma),\ldots,r_K^{(i)}(\hat{x}^\star_\gamma)\right]$, with $r^{(i)}_{\pi^{(i)}(k)}({\hat x}^\star_\gamma), k\in{\mathcal K}$, expressed as \eqref{eqn:new:XJ2} at the top of next page based on \eqref{eqn:new:XJ1}.
\begin{figure*}[!t]
\begin{align} \label{eqn:new:XJ2}
r^{(i)}_{\pi^{(i)}(k)}(\hat{x}^\star_\gamma)=
\frac{1}{T}\int_0^{\bar T}\log_2\left(\frac{\sigma^2+
\sum_{j=1}^{k}Ph_{\pi^{(i)}(j)}({\bar x}(t))}{\sigma^2+\sum_{j=1}^{k-1}Ph_{\pi^{(i)}(j)}({\bar x}(t))}\right)\textrm{d}t + \frac{\hat T}{T}\log_2\left(\frac{\sigma^2+
\sum_{j=1}^{k}Ph_{\pi^{(i)}(j)}(\hat{x}^\star_\gamma)}{\sigma^2+\sum_{j=1}^{k-1}Ph_{\pi^{(i)}(j)}(\hat{x}^\star_\gamma)}\right).
\end{align}
\end{figure*}
Let $\tau_{\gamma}^{(i)}$ denote the normalized time-sharing factor associated with the $\gamma$-th hovering location and the $i$-th decoding order, such that the UAV uses this strategy for a $\tau_\gamma^{(i)}$ portion of durations, where $\sum_{\gamma=1}^{\Gamma}\sum_{i=1}^{|{\cal I}|}\tau_\gamma^{(i)}=1$. Accordingly, finding the optimal time-sharing factors can be formulated as the following linear program (LP), which can be solved efficiently via convex optimization tools such as CVX \cite{CVX}.
\begin{align}
\textrm{(P1.3):}~&\max_{\{\tau^{(i)}_\gamma\ge 0\},R}~R \notag\\
&~~~~{\textrm{s.t.}}~\sum_{\gamma=1}^{\Gamma}\sum_{i=1}^{|{\cal I}|}\tau^{(i)}_\gamma r_{\pi^{(i)}(k)}^{(i)}(\hat{x}^\star_\gamma)\ge\alpha_kR,\ \forall k\in{\cal K}\notag\\
&~~~~~~~~~\sum_{\gamma=1}^{\Gamma}\sum_{i=1}^{|{\cal I}|}\tau_\gamma^{(i)}=1\notag.
\end{align}
Let $\{\tau^{(i)\star}_{\gamma}\}$ and $R^{\star}$ denote the optimal solution to problem (P1.3). Then the UAV needs to hover at each location $\hat x^\star_\gamma$ for duration $\sum_{i=1}^{|{\cal I}|}\tau_\gamma^{(i)\star}\hat T$. Accordingly, we partition the whole hovering period $\hat{\cal T}$ into $\Gamma$ sub-periods, denoted by $\hat{\cal T}_1,\ldots,\hat{\cal T}_\Gamma$, where $\hat{\cal T}_{\gamma}=\left(\sum_{j=1}^{\gamma-1}\sum_{i=1}^{|{\cal I}|}\tau_j^{(i)\star}\hat T,\sum_{j=1}^{\gamma}\sum_{i=1}^{|{\cal I}|}\tau_j^{(i)\star}\hat T\right],$ $\forall \gamma\in\{1,\ldots,\Gamma\}$. In this case, the optimal value (or the sum-rate capacity) of the primal problem (P1.2) is given by $R^{\star}$, and the corresponding optimal trajectory solution is given as
\begin{align}
\hat{x}^{\star}(t)=\hat{x}^\star_{\gamma},\ \forall t\in{\hat{\cal T}}_\gamma,\gamma\in\{1,\ldots,\Gamma\}.
\end{align}
Furthermore, in order to achieve the optimal communication rate $\mv r^{\star}$, the $i$-th decoding order ${\mv \pi}^{(i)}$ (together with the associated code rates) needs to be employed for a $\sum_{\gamma=1}^\Gamma \tau^{(i)\star}_\gamma$ portion of the whole duration $T$ in total. Therefore, problem (P1.2) is finally solved. The algorithm for optimally solving problem (P1.2) is summarized as Algorithm 1 in Table \ref{t2}.
\begin{table}[ht]
\begin{center}
\caption{\color{black}Algorithm 1 for Optimally Solving Problem (P1.2) under Given $x_\textrm{I}$ and $x_\textrm{F}$}\label{t2}
\hrule \vspace{1pt}
\begin{itemize}
\item[1)] {\bf Initialization}: Given an ellipsoid $\mathcal{E}^{(0)}({\mv \lambda},{\mv A})$ containing the optimal dual solution $\mv{\lambda}^\star=\left[\lambda_1^{\star},\ldots,\lambda_K^{\star}\right]$, where $\mv \lambda^{(0)}$ denotes the center point of ${\cal E}$ and the positive definite matrix ${\mv A}$ characterizes the size of ${\cal E}$, and set $n=0$.
\item[2)] {\bf Repeat:}
    \begin{itemize}
    \item[a)] Set permutation ${\mv \pi}$ such that $\lambda^{(n)}_{\pi(1)}\ge\ldots\ge\lambda^{(n)}_{\pi(K)}\ge0$, and then obtain $\hat{x}^{*}$ by solving problem \eqref{25.2} under given $\mv{\lambda}^{(n)}$ via a 1D exhaustive search over the region $[x_\textrm{I},x_\textrm{F}]$, and obtain ${\mv r}^{*}$ from \eqref{eqn:new:XJ1};
    \item[b)] Update the ellipsoid ${\cal E}^{(n+1)}$ via the ellipsoid method based on ${\cal E}^{(n)}$, and set $\mv \lambda^{(n+1)}$ as the center of ellipsoid $\mathcal{E}^{(n+1)}$;
    \item[c)] Set $n\leftarrow n+1$.
   \end{itemize}
\item[3)] {\bf Until} the stopping criteria for the ellipsoid method is met.
\item[4)] {\bf Set} ${{\mv\lambda}^{\star}}\leftarrow{{\mv \lambda}^{(n+1)}}$.
\item[5)] Obtain $\{\hat x^\star_\gamma\}_{\gamma=1}^\Gamma$ by solving problem \eqref{25.2} under ${{\mv\lambda}^{\star}}$, and construct the corresponding achievable rates ${\mv r}^{(i)}(\hat{x}_\gamma^\star)$.
\item[6)] Solve problem (P1.3) to obtain the optimal time-sharing factors for different hovering locations and decoding orders.
\end{itemize}\vspace{1pt}
\hrule
\end{center}
\end{table}
\vspace{1pt}
\begin{remark}
The optimal solution to problem (P1.2) reveals that to maximize the capacity region of the UAV-enabled MAC with speed-free trajectory, the UAV needs to hover above a finite number of ground locations with optimized durations, in order to balance the rate tradeoff among different users distributed on the ground. We refer to such a trajectory solution as the {\it multi-location-hovering}. Besides, the UAV also needs to properly time share among different decoding orders at each hovering location.
\end{remark}

\subsection{Optimal Solution to Problem (P1.1)}
Now, it remains to solve problem (P1.1). First, it is evident that the optimal solution of $\mv r^\star$ and $R^\star$ to problem (P1.2) is still optimal for problem (P1.1), for which the time-sharing among different decoding orders is needed. Next, we obtain the optimal trajectory solution $\{{x}^{\star}(t)\}$ to problem (P1.1) based on Lemma \ref{l2}, by combining the above (speed-free) multi-location-hovering solution $\{\hat{x}^{\star}(t)\}$ to problem (P1.2) and the maximum-speed flying trajectory $\{\bar{x}(t)\}$ in \eqref{9}. We thus have the following proposition, for which the proof is omitted for brevity.
\vspace{1pt}
\begin{proposition}\label{Proposition3.1}
The optimal trajectory solution $\{{x}^\star(t)\}$ to problem (P1.1) has the following SHF structure: The UAV unidirectionally flies from the initial location $x_{\rm I}$ to the final location $x_{\rm F}$, during which it successively hovers above locations $\hat{x}^\star_1,\ldots,\hat{x}^\star_{\Gamma}$, with durations $\sum_{i=1}^{|{\cal I}|}\tau_1^{(i)\star}\hat T,\ldots,$
$\sum_{i=1}^{|{\cal I}|}\tau_\Gamma^{(i)\star}\hat T$, respectively, and flies unidirectionally among them at the maximum speed $V_{\textrm{max}}$.
\end{proposition}

By combining $\{x^\star(t)\}$ in Proposition 3.1 together with $\mv r^\star$ and $R^\star$, problem (P1.1) is thus optimally solved. Notice that at the optimal solution to problem (P1.1), proper time-sharing among different decoding orders is required, similarly as for problem (P1.2).

Finally, by using the optimal solution to problem (P1.1), together with the 2D exhaustive search of $x_{\rm I}$ and $x_{\textrm{F}}$, the optimal solution to the original problem (P1) is obtained. It follows from Proposition \ref{Proposition3.1} that the optimal trajectory solution to problem (P1) also has the SHF structure.

{\it Complexity Analysis:} To analyze the implementation complexity of the algorithm for solving problem (P1), we first analyze that of Algorithm 1 for solving problem (P1.2). Suppose that for the exhaustive search, the same accuracy of $\varepsilon$ is used. Notice that the complexity of Algorithm 1 is dominated by the ellipsoid method in steps 2)-3) and the LP in step 5). In particular, the time complexity of step 2-a) is of order $K^2/\varepsilon$. Note that the ellipsoid method generally needs ${\mathcal{O}}\left(K^2\right)$ iterations to converge \cite{sb2}, and thus the time complexity of steps 2)-3) is ${\mathcal{O}}\left(K^4/\varepsilon\right)$. Furthermore, the time complexity for solving the LP in step 5) is ${\mathcal{O}}\left(\Gamma^2 |{\cal I}|^2 K\right)$ \cite{boyd}. Therefore, the time complexity of Algorithm 1 in Table \ref{t2} is ${\mathcal{O}}\left(K^4/\varepsilon+\Gamma^2 |{\cal I}|^2 K\right)$. In addition, in order to solve problem (P1), a 2D exhaustive search over $x_\textrm{I}$ and $x_\textrm{F}$ is required, with complexity of $\mathcal{O}(1/\varepsilon^2)$. As a result, the overall time complexity of the algorithm for solving problem (P1) is ${\mathcal{O}}\left(K^4/\varepsilon^3 + \Gamma^2 |{\cal I}|^2 K/\varepsilon^2\right)$.
\section{Rate Region Characterization under OMA}\label{V+}
In this section, we consider two OMA transmission schemes with FDMA and TDMA, in which different ground users communicate with the UAV over orthogonal frequency and time resources, respectively. For both schemes, we characterize the users' maximum achievable rate regions, by jointly optimizing the 1D UAV trajectory and wireless resource allocations.

First, we consider the FDMA case. At each time instant $t\in {\cal T}$, let $b_k(t)$ denote the (normalized) bandwidth allocated to user $k \in \mathcal K$. We have
\begin{align}
&b_k(t)\ge 0,\ \forall k\in {\cal K},t\in {\cal T},\label{b1}\\
&\sum_{k\in {\cal K}}b_k(t) = 1,\ \forall t\in {\cal T}\label{b2}.
\end{align}
The achievable rate (in bps/Hz) of user $k$ at time $t$ is given as $ b_k(t)\log_2\left(1+{Ph_k(x(t))}/{b_k(t)\sigma^2}\right)$, where $b_k(t)\sigma^2$ denotes the noise power at the UAV receiver over bandwidth $b_k(t)B$. Accordingly, under any given bandwidth allocation $\{b_k(t)\}$ and UAV trajectory $\{x(t)\}$, the achievable rate region by the $K$ users is given as
\begin{align}
&{\bar {\cal R}}_{\textrm{FDMA}}(\{b_k(t)\},\{x(t)\})=\left\{\mv r\in{\mathbb R}^{+}_K \bigg|r_k\leq\right.\notag\\
&~~~~\left.\frac{1}{T}\int^T_{0}b_k(t)\log_2\left(1+\frac{Ph_k(x(t))}{b_k(t)\sigma^2}\right)\textrm{d}t,\ \forall k\in{\cal K}\right\}.
\end{align}
Let $\mathcal{Y}_\textrm{FDMA}$ denote the feasible set of $\{b_k(t)\}$ specified by the constraints in \eqref{b1} and \eqref{b2}. The maximum achievable rate region is defined as
\begin{align}
{\cal R}_\textrm{FDMA}(V_{\textrm{max}},T)=~~~\bigcup_{\mathclap{\substack{  \{b_k(t)\}\in\mathcal{Y}_{\textrm{FDMA}},\\ \left\{x(t)\right\}\in{\mathcal X}}}}~~~{\bar{\cal R}}_\textrm{FDMA}(\{b_k(t)\},\left\{x(t)\right\}).
\end{align}
Similarly as the capacity-achieving NOMA case in Section II, we characterize the Pareto boundary of ${\cal R}_\textrm{FDMA}(V_{\textrm{max}},\!T)$ by using the rate-profile technique. By letting ${\mv \alpha}=[\alpha_1,\ldots,\alpha_K]$ denote a given rate-profile vector with $\alpha_k\geq 0, \forall k\in{\cal K}$, and $\sum_{k\in{\cal K}}\alpha_k=1$, we can characterize the Pareto boundary of ${\cal R}_\textrm{FDMA}(V_{\textrm{max}},T)$ under FDMA by solving the following problem:
\begin{align}
\textrm{(P2):}~&\max_{\{b_k(t)\},\left\{x(t)\right\},{\mv r},{R}}
~{R}\notag\\
&~~~~~~~~{\textrm{s.t.}}~\eqref{5},~\eqref{7},~\eqref{b1},~\eqref{b2}\notag\\
&~~~~~~~~~~~~~{\mv r}\in{{\mathcal {\bar R}}}_\textrm{FDMA}(\{b_k(t)\},\left\{x(t)\right\}).\label{32}
\end{align}

Next, we consider the TDMA case. At each time instant $t\in {\cal T}$, let $\rho_k(t)\in\{0,1\}, \forall k\in \mathcal K,$ denote a set of indicators for TDMA transmission. If user $k\in{\cal K}$ is scheduled to send its message to the UAV at time $t$, then we have $\rho_k(t)=1$; otherwise, we have $\rho_k(t)=0$. As only one user can be active in transmission at each time $t$, it follows that
\begin{align}
&\rho_k(t)\in\{0,1\},\ \forall k\in {\cal K},t\in {\cal T}, \label{b3}\\
&\sum_{k\in {\cal K}}\rho_k(t) = 1,\ \forall t\in {\cal T}\label{b4}.
\end{align}
Accordingly, the achievable rate (in bps/Hz) of user $k$ at time $t$ is given as $\rho_k(t)\log_2\left(1+{Ph_k(x(t))}/{\sigma^2}\right)$. Under any given $\{\rho_k(t)\}$ and UAV trajectory $\{x(t)\}$, the achievable rate region by the $K$ users is given as
\begin{align}
&\bar{\mathcal R}_{\textrm{TDMA}}(\{\rho_k(t)\},\{x(t)\})=\left\{\mv r\in{\mathbb R}^{+}_K \bigg|r_k\leq\right.\notag\\
&~~~~\left.\frac{1}{T}\int^T_0\!\rho_k(t)\log_2\!\left(1+\frac{Ph_k(x(t))}{\sigma^2}\right)\textrm{d}t,\ \forall k\in{\cal K}\right\}.
\end{align}
Let $\mathcal{Y}_{\textrm{TDMA}}$ denote the feasible set of $\{\rho_k(t)\}$ specified by the constraints in \eqref{b3} and \eqref{b4}. Then, the maximum achievable rate region is expressed as
\begin{align}
{\mathcal R}_\textrm{TDMA}(V_{\textrm{max}},T)=~~~\bigcup_{\mathclap{\substack{\{\rho_k(t)\}\in\mathcal{Y}_{\textrm{TDMA}},\\\left\{x(t)\right\}\in{\mathcal X}}}}~~~\bar{{\mathcal R}}_\textrm{TDMA}(\{\rho_k(t)\},\left\{x(t)\right\}).
\end{align}

By using the rate-profile technique with vector ${\mv \alpha}=[\alpha_1,\ldots,\alpha_K]$, we characterize the Pareto boundary of rate region ${\mathcal R}_\textrm{TDMA}(V_{\textrm{max}},T)$ by solving the following problem:
\begin{align}
\textrm{(P3):}~&\max_{\{\rho_k(t)\},\left\{x(t)\right\},\mv r,{R}}
~{R}\notag\\
&~~~~~~~~{\textrm{s.t.}}~\eqref{5},~\eqref{7},~\eqref{b3},~\eqref{b4}\notag\\
&~~~~~~~~~~~~~{\mv r}\in{{\mathcal {\bar R}}}_\textrm{TDMA}(\{\rho_k(t)\},\left\{x(t)\right\}).\label{61}
\end{align}

It is observed that problems (P2) and (P3) contain both UAV trajectory $\{x(t)\}$ and wireless resource allocation $\{b_k(t)\}$ and $\{\rho_k(t)\}$ as optimization variables, thus making them even more difficult to be solved than problem (P1). It is also worth noting that any feasible solution for problem (P3) in the TDMA case is also feasible for problem (P2) in the FDMA case; therefore, the achievable rate region by FDMA is generally larger than that by TDMA, i.e., ${\mathcal R}_\textrm{TDMA}(V_{\textrm{max}},T)\subseteq{\mathcal R}_\textrm{FDMA}(V_{\textrm{max}},T)$. In the following two subsections, we solve problems (P2) and (P3), respectively.

\subsection{Optimal Solution to Problem (P2) for FDMA}
This subsection solves problem (P2) for the FDMA case. Similar as for problem (P1) in the NOMA case, we consider the UAV trajectory $\{x(t)\}$ for problem (P2) to be unidirectional without loss of optimality. To obtain the optimal solution to problem (P2), we first consider the following problem (P2.1) under any given initial and final locations $x(0)=x_{\textrm{I}}$ and $x(T)=x_{\textrm{F}}$, with $w_1\le x_{\textrm{I}}\le x_{\textrm{F}}\le w_K$, and then use a 2D exhaustive search over $[w_1,w_K]\times[w_1,w_K]$ to find the optimal initial and final locations $x_{\textrm{I}}$ and $x_{\textrm{F}}$.
\begin{align}
\textrm{(P2.1):}~&\max_{\{b_k(t)\},\left\{x(t)\right\},{\mv r},{R}}
~{R}\notag\\
&~~~~~~~~{\textrm{s.t.}}
~\eqref{5},~\eqref{7},~\eqref{8},~\eqref{b1},~\eqref{b2},~\eqref{32}.\notag
\end{align}

In the following, we only need to focus on solving problem (P2.1) under any given $x_\textrm{I}$ and $x_\textrm{F}$. Similar as Lemma \ref{l2}, it can be verified that for any speed-constrained trajectory $\{{x}(t)\}_{t=0}^{{T}}$ (and the corresponding bandwidth allocation $\{b(t)\}_{t=0}^T$), we can equivalently construct a (fixed) maximum-speed flying trajectory $\{\bar{x}(t)\}_{t=0}^{\bar{T}}$ and an (optimizable) speed-free trajectory $\{\hat{x}(t)\}_{t=0}^{\hat{T}}$ (together with the corresponding bandwidth allocation), such that the combination of $\{\bar{x}(t)\}$ and $\{\hat{x}(t)\}$ can achieve the same rate region as that achieved by $\{x(t)\}$. For notational convenience, let $\{\bar b_k(t)\}$ and $\{\hat b_k(t)\}$ denote the corresponding bandwidth allocations associated with trajectory $\{{\bar x}(t)\}$ and $\{\hat{x}(t)\}$, respectively. Then we have $\bar{\mathcal R}_\textrm{FDMA}(\{{b}_k(t)\},\{x(t)\})=$ $\hat{\mathcal R}_\textrm{FDMA}(\{\bar{b}_k(t)\},\{\hat{b}_k(t)\},\{\bar x(t)\},\{\hat x(t)\})$, where $\hat{\mathcal R}_\textrm{FDMA}(\{\bar{b}_k(t)\},\{\hat{b}_k(t)\},\{\bar x(t)\},\{\hat x(t)\})$ is given as \eqref{33} at the top of this page.
\begin{figure*}
\begin{align} \label{33}
&\hat{\mathcal R}_\textrm{FDMA}(\{\bar{b}_k(t)\},\{\hat{b}_k(t)\},\{\bar x(t)\},\{\hat x(t)\})=\notag\\ &~~~~~~~~~~~~~~~~~~~~\left\{\mv r\in\mathbb{R}^+_K \bigg| r_k\leq\frac{1}{T}\left(\int^{\bar{T}}_{0}\bar{b}_k(t)\log_2\left(1+\frac{Ph_k(\bar{x}(t))}{\bar{b}_k(t)\sigma^2}\right)\textrm{d}t+\int^{\hat{T}}_{0}\hat{b}_k(t)\log_2\left(1+\frac{Ph_k(\hat{x}(t))}{\hat{b}_k(t)\sigma^2}\right)\textrm{d}t\right)\right\}.
\end{align}
\end{figure*}
By replacing $\bar{\mathcal R}_\textrm{FDMA}(\{{b}_k(t)\},\{x(t)\})$ as $\hat{\mathcal R}_\textrm{FDMA}(\{\bar{b}_k(t)\},\{\hat{b}_k(t)\},\{\bar x(t)\},\{\hat x(t)\})$, problem (P2.1) with {\it speed-constrained} trajectory design can be reformulated as the following problem (P2.2) that jointly optimizes the {\it speed-free} trajectory $\{\hat{x}(t)\}$ and the bandwidth allocation $\{\bar{b}_k(t)\}$ and $\{\hat{b}_k(t)\}$.
\begin{align}
\textrm{(P2.2):}&~\max_{\{\bar{b}_k(t)\},\{\hat{b}_k(t)\},\{\hat{x}(t)\},{\mv r},{R}}
~{R}\notag\\
{\textrm{s.t.}}&~\eqref{5},~\eqref{18}\notag\\
&~{\mv r}\in\hat{\mathcal R}_\textrm{FDMA}(\{\bar{b}_k(t)\},\{\hat{b}_k(t)\},\{\bar x(t)\},\{\hat x(t)\})\label{38}\\
&~\bar{b}_k(t)\ge 0,\ \forall k\in {\cal K}, t\in (0,\bar{T}]\label{40}\\
&~\hat{b}_k(t)\ge 0,\ \forall k\in {\cal K}, t\in (0,\hat{T}]\label{41}\\
&~\sum_{k\in {\cal K}}\bar{b}_k(t)= 1,\ \forall t\in (0,\bar{T}]\label{43}\\
&~\sum_{k\in {\cal K}}\hat{b}_k(t)=1,\ \forall t\in (0,\hat{T}]\label{42}.
\end{align}
Note that the strong duality holds between problem (P2.2) and its Lagrange dual problem. Therefore, we apply the Lagrange duality method to solve problem (P2.2).

Let $\mu_k\ge0$ denote the dual variable associated with the $k$-th constraint in \eqref{5}, $k\in{\cal K}$. Then the partial Lagrangian of problem (P2.2) is
\begin{align}
&{\cal L}_2(\{\mu_k\},\{\bar{b}_k(t)\},\{\hat{b}_k(t)\},\{\hat{x}(t)\},{\mv r},R)=\notag\\
&~~~~~~~~~~~~~~~~~~~~~~~~~~~~(1-\sum_{k\in{\cal K}}\mu_k\alpha_k)R+\sum_{k\in{\cal K}}\mu_kr_k.
\end{align}
Accordingly, the Lagrange dual function of problem (P2.2) is given as \eqref{55} at the top of next page.
\begin{figure*}
\begin{align} \label{55}
f_2\left(\{\mu_k\}\right)=&\max_{\{\bar{b}_k(t)\},\{\hat{b}_k(t)\},\{\hat{x}(t)\},{\mv r},R}~\sum_{k\in{\cal K}}{\cal L}_2(\{\mu_k\},\{\bar{b}_k(t)\},\{\hat{b}_k(t)\},\{\hat{x}(t)\},{\mv r},R)\\
&~~~~~~~~~~~~~{\textrm{s.t.}}~\eqref{18},~\eqref{38},~\eqref{41},~\eqref{40},~\eqref{42},~\eqref{43}.\notag
\end{align}
\end{figure*}
Note that $f_2\left(\{\mu_k\}\right)$ is bounded from above only when $\sum\limits_{k\in\cal K}\mu_k \alpha_k=1$. Then the dual problem of problem (P2.2) is given by
\begin{align}
\textrm{(D2.2):}~&\min_{\{\mu_k \ge 0\}}~f_2\left(\{\mu_k\}\right)\notag\\
&~~{\textrm{s.t.}}
~\sum\limits_{k\in{\cal K}}\mu_k \alpha_k=1.\label{46}
\end{align}

As the strong duality holds between problems (P2.2) and (D2.2), we solve problem (P2.2) by equivalently solving problem (D2.2). First, we obtain the dual function $f_2\left(\{\mu_k\}\right)$ by solving problem \eqref{55} under any given $\{\mu_k\}$ that satisfies \eqref{46} and $\mu_k \ge 0, \forall k\in {\cal K}$. In this case, problem \eqref{55} is re-expressed as
\begin{align}\label{57}
&\max_{\substack{\{\bar{b}_k(t)\},\{\hat{b}_k(t)\},\\\{\hat{x}(t)\}}} \sum\limits_{k\in{\cal K}}\frac{\mu_k}{T}\left(\int^{\bar{T}}_{0}\!\bar{b}_k(t)\log_2\left(1+\frac{Ph_k(\bar{x}(t))}{\bar{b}_k(t)\sigma^2}\!\right)\!\textrm{d}t~+\!\right.\notag\\
&~~~~~~~~~~~~~~~\left.\int^{\hat{T}}_{0}\!\hat{b}_k(t)\log_2\left(1+\frac{Ph_k(\hat{x}(t))}{\hat{b}_k(t)\sigma^2}\right)\!\textrm{d}t\!\right)\\
&~~~~~~{\textrm{s.t.}}~\eqref{18},~\eqref{41},~\eqref{40},~\eqref{42},~\eqref{43}.\notag
\end{align}

Note that problem \eqref{57} can be decomposed into two sets of subproblems as follows, problem \eqref{58} for optimizing $\{{\bar b}_k(t)\}$ at any time $t\in(0,\bar{T]}$ under given UAV location ${\bar x}(t) = x_\textrm{I} + V_\textrm{max} t$, and problem \eqref{59} for jointly optimizing $\{{\hat b}_k(t)\}$ and $\hat x(t)$ at any time $t\in(0,\hat{T]}$.
\begin{align}\label{58}
&~\max_{\{\bar{b}_k(t)\ge 0\}}~\sum\limits_{k\in{\cal K}}\frac{\mu_k}{T}\bar{b}_k(t)\log_2\left(1+\frac{Ph_k(\bar{x}(t))}{\bar{b}_k(t)\sigma^2}\right)\\
&~~~~{\textrm{s.t.}}~\sum_{k\in \cal K} {\bar b}_k(t)=1.\notag\\
&\max_{\substack{\{\hat{b}_k(t)\ge 0\},\\x_\textrm{I} \le \hat{x}(t)\le x_\textrm{F}}}~\sum\limits_{k\in{\cal K}}\frac{\mu_k}{T}\hat{b}_k(t)\log_2\left(1+\frac{Ph_k(\hat{x}(t))}{\hat{b}_k(t)\sigma^2}\right)\label{59}\\
&~~~~{\textrm{s.t.}}~\sum_{k\in \cal K} {\hat b}_k(t)=1.\notag
\end{align}

To facilitate the solution of problems \eqref{58} and \eqref{59}, we define that under any given dual variables $\mu_k$'s and UAV location $x$, the correspondingly achieved maximum weighted sum-rate is given as the following function $g_1(\{\mu_k\},x)$, which is obtained by optimizing the bandwidth allocation $b_k$'s.
\begin{align}
g_1(\{\mu_k\},x)=&\max_{\{{b}_k\ge 0\}}~\sum\limits_{k\in{\cal K}}{\mu_k}{b}_k\log_2\left(1\!+\!\frac{Ph_k(x)}{{b}_k\sigma^2}\right)\label{60}\\
&~~{\textrm{s.t.}}~\sum_{k\in {\cal K}}b_k = 1\label{78}.
\end{align}
As the problem in \eqref{60} is convex, we use the Karush-Kuhn-Tucker (KKT) conditions \cite{boyd} to obtain its optimal solution, which is given as
\begin{align}
{b}_k^{(\{\mu_k\},x)}\!=\!\frac{-\mathcal{W}\left(-{\exp\left(-\!\left(\frac{\eta  T\ln2}{\mu_k}\!+\!1\right)\right)}\right)Ph_k(x)}{\left(1\!+\!\mathcal{W}\left(-{\exp\left(-\!\left(\frac{\eta  T\ln2}{\mu_k}\!+\!1\right)\right)}\right)\right)\sigma^2},\ \forall k\in {\cal K}.\label{82}
\end{align}
Here, $\mathcal{W}(\cdot)$ is the Lambert $\mathcal{W}$ function with $\mathcal{W}(x) e^{\mathcal{W}(x)} = x$ \cite{W}, and $\eta \ge 0$ denotes the optimal dual variable associated with constraint \eqref{78} that can be obtained via a bisection search based on $\sum_{k\in \cal K} b_k^{(\{\mu_k\},x)} = 1$.

Now, we solve problems \eqref{58} and \eqref{59} based on problem \eqref{60} and \eqref{82}. By comparing problems \eqref{58} and \eqref{60}, it is evident that the optimal solution of $\{\bar{b}_k(t)\}$ to problem \eqref{58} is given as ${\bar b}_k^*(t) = b_k^{(\{\mu_k\},\bar{x}(t))}, \forall k\in {\cal K}, t\in (0,\bar T]$, with $\{b_k^{(\{\mu_k\},{x})}\}$ defined in \eqref{82}. As for problem \eqref{59}, it is evident that under any given ${\hat x}(t) = x$, the optimal bandwidth allocation is ${\hat b}_k^*(t)=b_k^{(\{\mu_k\},{x})}, \forall k\in \cal K$, and the accordingly achieved objective value is $g_1(\{\mu_k\},x)/T$. In this case, we adopt a 1D exhaustive search over $[x_\textrm{I},x_\textrm{F}]$ to find the optimal UAV location solution $\hat{x}^*(t)$ to problem \eqref{59} as
\begin{align}
\hat x^*(t) = \hat{x}^* \triangleq  \arg~\max\limits_{\mathclap{x_\textrm{I} \le x \le x_\textrm{F}}}~g_1(\{\mu_k\},x),\ \forall t\in (0,\hat T].\label{10}
\end{align}
Accordingly, the optimal bandwidth allocation solution to problem \eqref{59} is given by
\begin{align}
{\hat b}_k^*(t) = b_k^{(\{\mu_k\},\hat{x}^*(t))},\ \forall k\in {\cal K}, t\in (0,\hat T].
\end{align}
With problems \eqref{58} and \eqref{59} solved, the dual function $f_2({\{\mu_k\}})$ is thus obtained.

Next, with $f_2({\{\mu_k\}})$ at hand, we obtain the optimal dual variable $\{\mu_k^\star\}$ by solving problem (D2.2) via standard subgradient-based methods such as the ellipsoid method, for which the details are omitted. After that, we still need an additional step to reconstruct the optimal primal solution to problem (P2.2), as the optimal solution to problem \eqref{59} may not be unique in general. In particular, under the optimal dual variable $\{\mu_k^\star\}$, suppose that there are $\Upsilon \ge 1$ UAV location solutions to problem \eqref{59}, denoted by $\{\hat{x}^\star_\upsilon\}_{\upsilon=1}^\Upsilon$, with $\hat{x}^\star_1\le\ldots\le\hat{x}^\star_\Upsilon$. In this case, we need to time-share the $\Upsilon$ solutions by allowing the UAV to hover at each of the $\Upsilon$ locations for a certain duration that needs to be optimized.

We denote $\left\{{\tilde r}_k^\textrm{FDMA}(\hat{x}^\star_\upsilon)\right\}$ as the users' rate tuples that is achieved when the UAV stays at location $\hat{x}^\star_\upsilon$, with the corresponding optimal bandwidth allocation as $\{b_k^{(\{\mu_k^\star\},\hat{x}^\star_\upsilon)}\}$, i.e.,
\begin{align}
{\tilde r}_k^\textrm{FDMA}(\hat{x}^\star_\upsilon)\!=&\frac{1}{T}\!\int^{\bar{T}}_{0}\!{b}_k^{(\{\mu^\star_k\},\bar{x}(t))}\log_2\!\left(1\!+\!\frac{Ph_k(\bar{x}(t))}{{b}_k^{(\{\mu^\star_k\},\bar{x}(t))}\sigma^2}\right)\textrm{d}t~+\notag\\ &\frac{\hat{T}}{T}{b}_k^{(\{\mu^\star_k\},\hat{x}^\star_\upsilon)}\log_2\left(1\!+\!\frac{Ph_k(\hat{x}^\star_\upsilon)}{{b}_k^{(\{\mu^\star_k\},\hat{x}^\star_\upsilon)}\sigma^2}\right), \ \forall k\in{\cal K}.\label{63}
\end{align}
Let $\{\kappa_\upsilon\ge0\}$ denote the normalized time-sharing factors among the $\Upsilon$ UAV hovering locations, where $\sum_{\upsilon=1}^{\Upsilon}\kappa_\upsilon=1$. We can thus obtain the optimal time-sharing factors $\{\kappa_\upsilon^\star\}$ as
\begin{align}
(\{\kappa_\upsilon^\star\},R^\star) = \arg~&\max_{\mathclap{\{\kappa_\upsilon\ge 0\},R}}~~R\label{84}\\
&~{\textrm{s.t.}}~\sum_{\upsilon=1}^{\Upsilon}\kappa_\upsilon {\tilde r}_k^\textrm{FDMA}(\hat{x}^\star_\upsilon)\ge\alpha_kR,\ \forall k\in{\cal K}\notag\\
&~~~~~~\sum_{\upsilon=1}^{\Upsilon}\kappa_\upsilon=1\notag.
\end{align}
Accordingly, the achieved rate tuple ${\mv r}^\star= \left[r_1^\star,\ldots, r_K^\star\right]$ is obtained as $r_k^\star = \alpha_k R^\star, \forall k\in \cal K$. Therefore, the optimal solution to problem (P2.2) is obtained as follows.

\vspace{1pt}
\begin{proposition}
The optimal speed-free UAV trajectory solution $\{\hat x^\star(t)\}$ to problem (P2.2) follows the multi-location-hovering structure: The UAV hovers above the $\Upsilon$ locations, $\hat x_1^\star,\ldots,\hat x_\Upsilon^\star$, each for duration $\kappa_\upsilon^\star \hat T$, i.e.,
\begin{align}
\hat{x}^{\star}(t)=\hat{x}^\star_{\upsilon},\ \forall t\in{\hat{\cal T}}_\upsilon,\upsilon\in\{1,\ldots,\Upsilon\},
\end{align}
where ${\hat{\cal T}}_\upsilon=\left(\sum_{j=1}^{\upsilon-1}\kappa^\star_j\hat{T},\sum_{j=1}^{\upsilon}\kappa^\star_j\hat{T}\right],\upsilon\in\{1,\ldots,\Upsilon\},$ denotes the hovering period at location $\hat{x}^\star_\upsilon$. The corresponding bandwidth allocations at each hovering location $\hat{x}^\star_{\upsilon}$ and maximum-speed flying location $\bar x(t)$ are given as ${\hat b}_k^\star(t) = b_k^{(\{\mu_k^\star\},\hat{x}^\star_\upsilon)}, \forall k\in {\cal K}, t\in{\hat{\cal T}}_\upsilon,$ and ${\bar b}_k^\star(t) = b_k^{(\{\mu^\star_k\},\bar{x}(t))}, \forall k\in {\cal K}, t\in (0,\bar T]$, respectively. Accordingly, the optimal achievable rates for the $K$ users and sum-rate are given by ${\mv r}^\star$ and $R^\star$, respectively.
\end{proposition}

Furthermore, by combining the speed-free trajectory $\{\hat x^\star(t)\}$  to problem (P2.2) and the maximum-speed flying trajectory $\{\bar x(t)\}$, together with the corresponding bandwidth allocations, the optimal solution to problem (P2.1) is obtained in the following proposition, for which the proof is omitted for brevity.
\vspace{1pt}
\begin{proposition}
The optimal UAV trajectory solution $\{{x}^\star(t)\}$ to problem (P2.1) follows the SHF structure: The UAV unidirectionally flies from the initial location $x_\textrm{I}$ to the final location $x_\textrm{F}$, during which it successively hovers above locations $\hat{x}_1^\star,\ldots,\hat{x}^\star_\Upsilon$, with durations $\kappa_1^\star\hat{T},\ldots,\kappa_\Upsilon^\star\hat{T}$, respectively, and flies among them at the maximum speed $V_\textrm{max}$. The bandwidth allocation is given as ${b}_k^\star(t)= b_k^{( \{\mu_k^\star\},x^\star(t))}, \forall t\in{\cal T}$, with $\{b_k^{(\{\mu_k\},{x})}\}$ defined in \eqref{82}. The correspondingly achieved rates are given as ${\mv r}^\star$ and $R^\star$, which are identical to those to problem (P2.2).
\end{proposition}

Therefore, the optimal solution to problem (P2.1) is finally obtained. By using the optimal solution to problem (P2.1) together with a 2D exhaustive search over $x_{\rm I}$ and $x_{\textrm{F}}$ with $w_1 \le x_\textrm{I} \le x_\textrm{F} \le w_K$, the optimal solution to the original problem (P2) is finally obtained.

\subsection{Optimal Solution to Problem (P3) for TDMA}
This subsection solves problem (P3) for the TDMA case. First, it can be easily verified that the optimal UAV trajectory solution to problem (P3) follows a unidirectional SHF structure with hovering locations exactly above users. This is due to the fact that by hovering exactly above the correspondingly communicating user, the UAV can achieve the lowest path loss and maximum data rate, under our considered probabilistic LoS channels. In this case, let $x_\textrm{I}$ and $x_\textrm{F}$ denote the initial and final locations, respectively. Then, similarly as for problems (P1) and (P2), the original speed-constrained trajectory $\{{x}(t)\}_{t=0}^{{T}}$ (with time allocation $\{\rho_k(t)\}$) for problem (P3) is equivalent to the combination of a fixed maximum-speed flying trajectory $\{\bar{x}(t)\}_{t=0}^{\bar{T}}$ (with time allocation $\{\bar{\rho}_k(t)\}$) and an optimizable speed-free trajectory $\{\hat{x}(t)\}_{t=0}^{\hat{T}}$ (with time allocation $\{\hat{\rho}_k(t)\}$), in terms of their achieved rate regions. More specifically, it is evident that the speed-free trajectory $\{{\hat x}(t)\}_{t=0}^{\hat T}$ has the multi-location-hovering structure with $w_k$'s being hovering locations. In this case, the speed-free trajectory $\{{\hat x}(t)\}_{t=0}^{\hat T}$ is only dependent on the hovering durations, denoted by $\hat{\tau}_k \ge 0$ for hovering location $w_k, k\in {\cal K}$, i.e., we have $\hat x(t) = w_k, \forall k\in {\cal K}, t\in \left(\sum_{i=1}^{k-1}\hat\tau_i,\sum_{i=1}^{k}\hat\tau_i\right]$.

Next, consider the time allocation under the multi-location-hovering speed-free trajectory. When the UAV hovers above each user $k\in{\cal K}$, it will communicate with that user, i.e., it follows that
\begin{align}
\hat{\rho}_k(t)=1,\hat \rho_j(t)=0,\ \forall j\in{\cal K}, j\neq k,t \in\left(\sum_{i=1}^{k-1}\hat{\tau}_i,\sum_{i=1}^{k}\hat{\tau}_i\right].
\end{align}
Here, notice that we have $\hat{\tau}_k = 0$, if $w_k < x_\textrm{I}$ or $w_k > x_\textrm{F}$. For notational convenience, we define user indexes $k_\textrm{I}$ and $k_\textrm{F}$ (with $1\le k_\textrm{I} \le k_\textrm{F} \le K$) and set ${\hat {\cal K}}\triangleq\{k_\textrm{I},\ldots,k_\textrm{F}\}$, such that $w_{k_\textrm{I}-1} < x_\textrm{I} \le w_{k_\textrm{I}}$ and $w_{k_\textrm{F}} \le x_\textrm{F} < w_{k_\textrm{F} + 1}$, with $w_0 <w_1$ and $w_{K+1} > w_{K}$ defined.

Now, by combining the above trajectories and time allocations, we have $\bar{\mathcal R}_{\textrm{TDMA}}(\{\rho_k(t)\},\{x(t)\})=\hat{\mathcal R}_\textrm{TDMA}(\{\bar{\rho}_k(t)\},\{\hat{\tau}_k\},\{\bar x(t)\})$, where $\hat{\mathcal R}_\textrm{TDMA}(\{\bar{\rho}_k(t)\},\{\hat{\tau}_k\},\{\bar x(t)\})$ is given as \eqref{56} at the top of next page.
\begin{figure*}
\begin{align} \label{56}
&\hat{\mathcal R}_\textrm{TDMA}(\{\bar{\rho}_k(t)\},\{\hat{\tau}_k\},\{\bar x(t)\})=\left\{\mv r\in\mathbb{R}^+_K \bigg| r_k\leq\frac{1}{T}\!\int^{\bar{T}}_{0}\!\bar{\rho}_k(t)\log_2\!\left(1\!+\!\frac{Ph_k(\bar{x}(t))}{{\sigma^2}}\right)\textrm{d}t\!+\!\frac{\hat{\tau}_k}{T}\log_2\!\left(1\!+\!\frac{Ph_k(w_k)}{\sigma^2}\right)\!,\ \forall k\in{\cal K}\right\}.
\end{align}
\end{figure*}
Therefore, by replacing $\bar{\mathcal R}_\textrm{TDMA}(\{{\rho}_k(t)\},\{x(t)\})$ as $\hat{\mathcal R}_\textrm{TDMA}(\{\bar{\rho}_k(t)\},\{\hat{\tau}_k\},\{\bar x(t)\})$ in problem (P3) and considering given $x_\textrm{I}$ and $x_\textrm{F}$ (thus given $\{{\bar x}(t)\}$), we have the following problem (P3.1) that jointly optimizes the time allocation $\{\bar{\rho}_k(t)\}$ under the fixed maximum-speed flying trajectory and hovering duration $\{\hat{\tau}_k\}$ under given hovering locations above each user.
\begin{align}
\textrm{(P3.1):}~&{\max_{\{\bar{\rho}_k(t)\},\{\hat{\tau}_k\ge 0\},{\mv r},{R}}
~{R}}\notag\\
&~~~~~~~~{\textrm{s.t.}}~{\eqref{5}}\notag\\
&~~~~~~~~~~~~~{\mv r}\in\hat{\mathcal R}_\textrm{TDMA}(\{\bar{\rho}_k(t)\},\{\hat{\tau}_k\},\{\bar x(t)\})\label{64}\\
&~~~~~~~~~~~~~\bar{\rho}_k(t)\in\{0,1\},\ \forall k\in {\cal K}, t\in (0,\bar{T}]\label{66}\\
&~~~~~~~~~~~~~\sum_{k\in {\cal K}}\bar{\rho}_k(t) = 1,\ \forall t\in (0,\bar{T}]\label{68}\\
&~~~~~~~~~~~~~\hat{\tau}_k=0,\ \forall k\in{\cal K}\setminus\hat{\cal K}\label{92}\\
&~~~~~~~~~~~~~\sum\limits_{k\in\hat{\cal K}}\hat{\tau}_k=\hat{T}.\label{93}
\end{align}
Therefore, we can solve the rate region characterization problem (P3) for TDMA, by first solving problem (P3.1) under any given initial and final locations $x_\textrm{I}$ and $x_\textrm{F}$, and then searching over $[w_1,w_K]\times[w_1,w_K]$ to find the optimal initial and final locations $x_\textrm{I}$ and $x_\textrm{F}$.

It thus remains to solve problem (P3.1) under given initial and final locations $x_\textrm{I}$ and $x_\textrm{F}$. Note that the strong duality holds between problem (P3.1) and its Lagrange dual problem, we solve problem (P3.1) via the Lagrange duality method. The optimal solution is presented as follows, for which the detailed derivations are similar to those for solving problems (P1.2) and (P2.2) and thus omitted.

In particular, let $\{\nu_k^\star\}$ denote the optimal dual variables associated with the $K$ constraints in \eqref{5}, which can be obtained by solving the dual problem of problem (P3.1) via the ellipsoid method. Accordingly, under any given UAV location $x$, we define $k^{( \{\nu^\star_j\},x)}$ as the user index that achieves the maximum weighted rate, i.e.,
\begin{align}
k^{(\{\nu^\star_j\},x)} = \arg\max\limits_{j\in\cal K}~{\nu_j^\star}\log_2\left(1+\frac{Ph_j(x)}{\sigma^2}\right).\label{94}
\end{align}
\vspace{1pt}
\begin{proposition}\label{P4.3}
The optimal time allocation solution of $\{\bar{\rho}_k(t)\}$ to problem (P3.1) is given as
\begin{align}
\bar{\rho}^\star_k(t)=\left\{
\begin{array}{rcl}
1,&&{\textrm{if}~k=k^{(\{\nu^\star_j\},\bar{x}(t))}},\\
0,&&{\textrm{if}~k\ne k^{(\{\nu^\star_j\},\bar{x}(t))}},
\end{array} \right.
\ \forall k\in{\cal K},t\in (0,\bar T].\label{90}
\end{align}
Accordingly, the optimal $\{\hat{\tau}^\star_k\}$, ${\mv r}^\star = [r_1^\star,\ldots,r_K^\star]$, and $R^\star$ can be obtained by solving the following LP via, e.g., CVX \cite{CVX}.
\begin{align}
&\max_{{\{\hat{\tau}_k\ge0\}},{\mv r},R}~R\notag\\
&~~~~~{\textrm{s.t.}}~\eqref{5},~\eqref{92},~\eqref{93}\notag\\
&~~~~~~~~~~r_k\le\frac{1}{T}\int^{\bar{T}}_{0}\!\bar{\rho}^\star_k(t)\log_2\left(1+\frac{Ph_k(\bar{x}(t))}{{\sigma^2}}\right)\textrm{d}t~+\notag\\
&~~~~~~~~~~~~~~~~{\frac{\hat{\tau}_k}{T}}\log_2\!\left(1\!+\!\frac{Ph_k(w_k)}{\sigma^2}\right),\ \forall k\in{\cal K}.\notag
\end{align}
\end{proposition}

Therefore, the optimal solution to problem (P3.1) is finally obtained. By using Proposition \ref{P4.3} together with a 2D exhaustive search over $w_1 \le x_\textrm{I} \le x_\textrm{F} \le w_K$, we can find the optimal initial and final locations as $x_\textrm{I}^\star$ and $x_\textrm{F} ^\star$. Let $k_\textrm{I}^\star$ and $k_\textrm{F}^\star$ as the corresponding $k_\textrm{I}$ and $k_\textrm{F}$. In this case, by further combining the obtained multi-location-hovering trajectory $\{\hat{x}(t)\}$ and the maximum-speed flying trajectory $\{\bar{x}(t)\}$, together with the corresponding time allocations, we finally have the optimal solution to the original problem (P3) in the following proposition.
\vspace{1pt}
\begin{proposition}\label{c2}
The optimal UAV trajectory solution $\{{x}^\star(t)\}$ to problem (P3) follows the SHF structure: The UAV unidirectionally flies from the optimal initial location $x^\star_\textrm{I}$ to the optimal final location $x^\star_\textrm{F}$, during which it successively hovers above locations $w_{k^\star_\textrm{I}},\ldots,w_{k^\star_\textrm{F}}$, with durations $\hat{\tau}_{k^\star_\textrm{I}}^\star,\ldots,\hat{\tau}_{k^\star_\textrm{F}}^\star$, respectively, and flies among them at the maximum speed $V_\textrm{max}$. The optimal time allocation is given as
\begin{align}
{\rho}^\star_k(t)=\left\{
\begin{array}{rcl}
1,&&{\textrm{if}~k=k^{(\{\nu^\star_j\},{x}^\star(t))}},\\
0,&&{\textrm{if}~k\ne k^{(\{\nu^\star_j\},{x}^\star(t))}},
\end{array} \right.
\ \forall k\in{\cal K}, t\in\cal T,
\end{align}
with $k^{(\{\nu^\star_j\},x)}$ defined in \eqref{94}.
The correspondingly achieved rates are given as ${\mv r}^\star$ and $R^\star$, which are identical to those for problem (P3.1) under optimal $x^\star_\textrm{I}$ and $x^\star_\textrm{F}$.
\end{proposition}
\vspace{1pt}
\begin{remark}
Based on Proposition \ref{c2}, it is evident that the UAV only hovers exactly above (some or all of) the users. As the UAV flies unidirectionally over time, the weighted rate $\nu_k^\star \log_2 (1+{Ph_k(x^\star(t))}/{\sigma^2})$ for each user $k \in \cal K$ first increases (when the UAV approaches) and then decreases (when the UAV flies away). Therefore, it can be easily shown that these users communicate with the UAV in a sequential transmission manner, i.e., user 1 first, followed by user 2, user 3, and so on, until user $K$. Furthermore, each user $k$ and the subsequent user $k+1$ switch their transmission at location $x$ such that ${\nu_k}\log_2\left(1+{Ph_k(x)}/{\sigma^2}\right) ={\nu_{k+1}}\log_2\left(1+{Ph_{k+1}(x)}/{\sigma^2}\right),k\in\{1,\ldots,K-1\}$.
\end{remark}
\vspace{1pt}
\begin{remark}\label{r4}
It is interesting to compare the optimal solutions to problems (P1), (P2), and (P3) for NOMA, FDMA, and TDMA, respectively. First, all the three trajectory solutions are observed to have the SHF structure. For NOMA and FDMA, the hovering locations are generally above middle points among users due to their simultaneous transmission; while for TDMA, each hovering location is exactly above one user for individual transmission. Next, wireless resources should be properly allocated jointly with the UAV trajectory design. For NOMA, it is crucial to find the optimal time sharing among different decoding orders; for FDMA, it is critical to design the optimal frequency allocation among users based on the UAV's location over time; for TDMA, different users communicate with the UAV in a sequential manner, and it is important to find the switching points for different users' communications.
\end{remark}
\section{Numerical Results}\label{sec:V}
In this section, we present numerical results to validate the performance of our proposed UAV trajectory designs as compared with the following two benchmark schemes.

\subsubsection{Successive Hovering Above Users}
The UAV successively visits the $K$ users with the initial and final locations being $x_\textrm{I}=w_1$ to $x_\textrm{F}=w_K$. The UAV flies at the maximum speed $V_\textrm{max}$, and only hovers above these users at $w_1,\ldots, w_K$. The hovering durations are optimized jointly with the wireless resource allocations, i.e., the time-sharing factors for decoding orders in NOMA, and the frequency allocation and the time allocation for FDMA and TDMA, respectively. Note that this design is only implementable when $T\ge{(w_K-w_1)}/{V_{\textrm{max}}}$ in order to visit all users. Also note that for the TDMA scheme, this scheme is identical to the optimal solution to problem (P3) when $T \ge {(w_K-w_1)}/{V_{\textrm{max}}}$.
\subsubsection{Static Hovering}
The UAV hovers at one single optimized location $x_{\textrm{H}}$ over the whole communication period without any movement. In this case, the capacity/rate region characterization problem under any given UAV location corresponds to a conventional wireless resource allocation problem for all of the three schemes with NOMA, FDMA, and TDMA. By solving the wireless resource allocation problem together with a 1D search of the UAV hovering location, we can obtain the corresponding capacity or rate regions.

In the simulation, we set the UAV's flight altitude as $H=250$~m, maximum speed as $V_{\textrm{max}}=20$ m/s, and noise power as $\sigma^2=-100$ dBm. We assume that the parameters of the probabilistic LoS channel model in \eqref{3} are set as $C=10$, $D=0.6$, and $\xi=0.2$. We also set the reference channel power gain as $\beta_0=-30$ dB, and the transmit power at each user as $P=30$ dBm.

\begin{figure}[t]
\centering
\includegraphics[width=8cm]{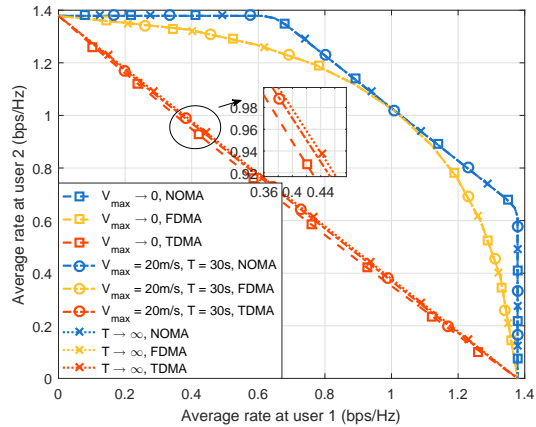}
\caption{Capacity and rate regions in the case with $K=2$ users and $D=100$ m.}\label{f2}
\end{figure}

First, we consider the case with $K=2$ users, where $D > 0$ denotes their distance. Fig. \ref{f2} shows the capacity and rate regions under NOMA, FDMA, and TDMA, in the case with $D = 100$ m. It is observed that the capacity region ${\cal C}(V_{\textrm{max}},T)$ under NOMA is significantly larger than the rate regions under FDMA/TDMA. It is also observed that FDMA considerably outperforms TDMA. This is due to the fact that for the rate region characterization problem (P3) under TDMA, any feasible solution is also feasible to problem (P2) under FDMA. The reverse, however, is not true. Therefore, the optimal value achieved by problem (P2) is always no smaller than that by problem (P3). Moreover, it is also observed that the capacity/rate regions under NOMA/FDMA are convex. This is due to the fact that with relatively short distance of $D=100$ m, the UAV should hover at one single location during the whole communication period. This shows that in this case with NOMA/FDMA, the UAV's mobility cannot increase the capacity/rate regions. By contrast, it is observed that when $V_\textrm{max}$ and $T$ are both finite, the rate region by TDMA is non-convex, as the UAV needs to hover above different users to efficiently collect information from them. In this case, the UAV's mobility is beneficial in increasing the rate region for TDMA.

\begin{figure}[t]
\centering
\includegraphics[width=8cm]{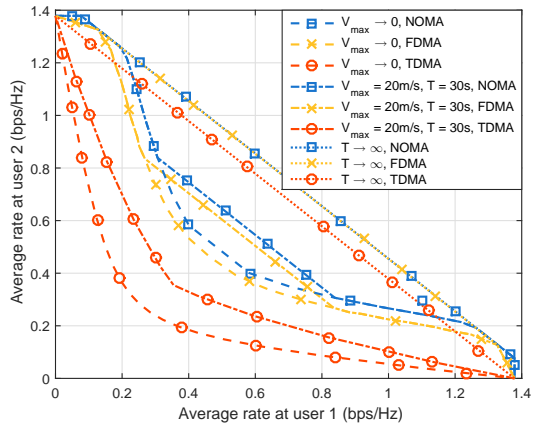}
\caption{Capacity and rate regions in the case with $K=2$ users and $D=800$ m.}\label{f3}
\end{figure}

Fig. \ref{f3} shows the capacity/rate region in the case with $D=800$ m. It is observed that under finite $T$ and $V_\textrm{max}$, the capacity and rate regions under the three schemes of NOMA, FDMA, and TDMA are all non-convex, as the UAV needs to visit different locations for collecting information. In this case, the UAV's mobility can significantly enlarge the corresponding capacity and rate regions. Furthermore, it is observed that FDMA achieves a rate region that is close to the capacity region by NOMA, and significantly outperforms TDMA.

\begin{figure}[t]
\centering
\includegraphics[width=8cm]{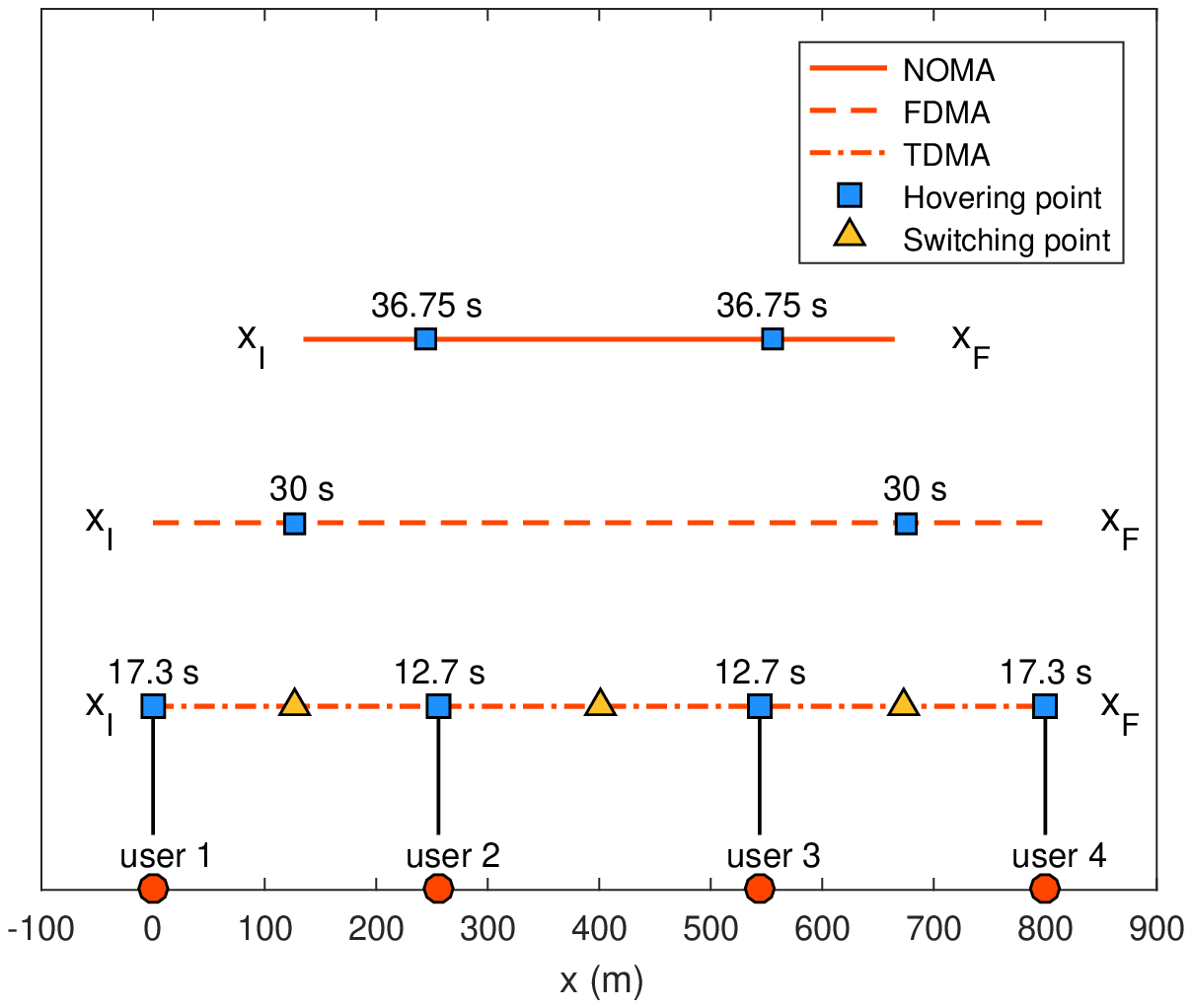}
\caption{Optimized UAV trajectories in the case with uniformly distributed users.}\label{f5}
\vspace{0.41cm}
\includegraphics[width=8cm]{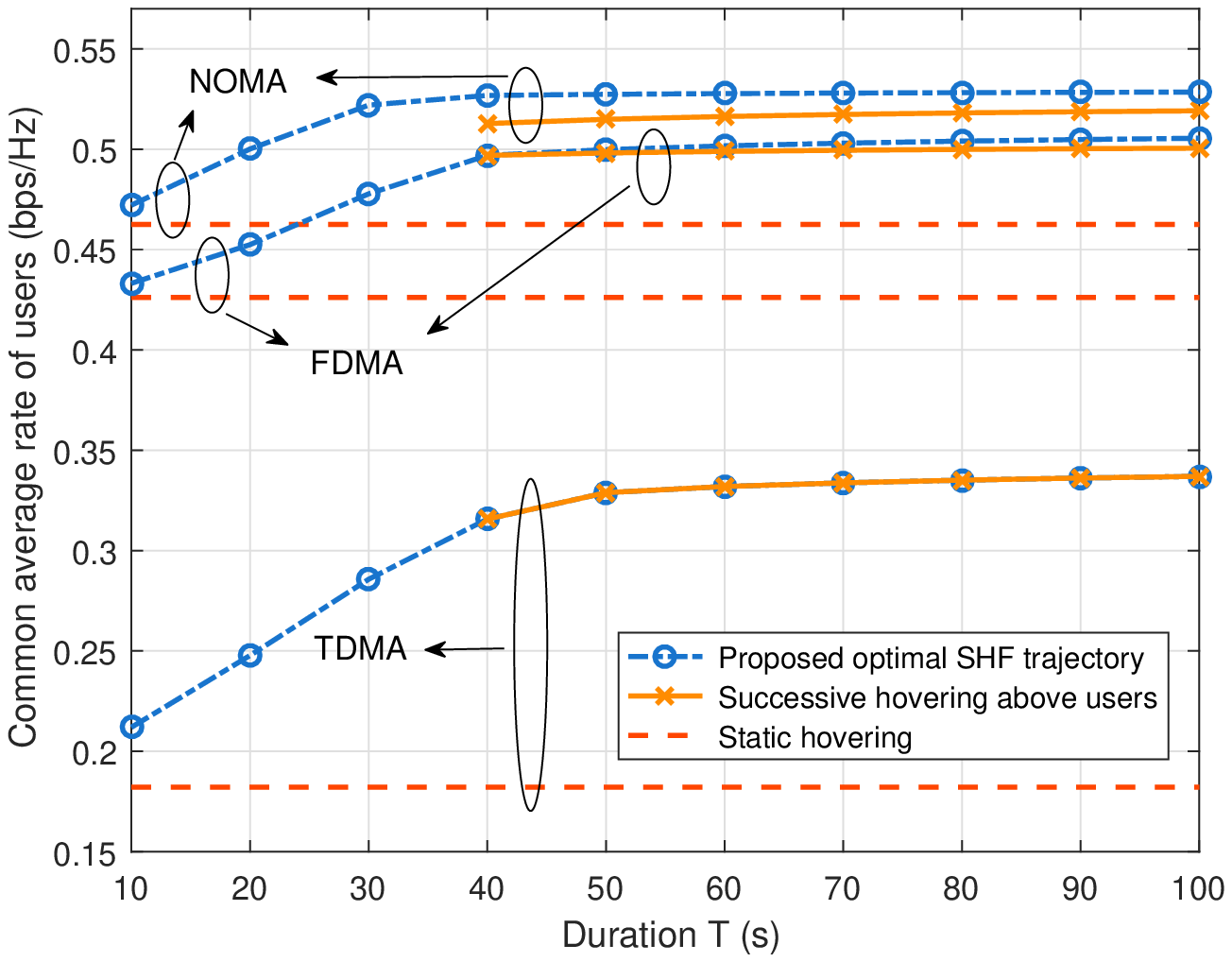}
\caption{Common average rate versus duration $T$ in the case with uniformly distributed users.}\label{f6}
\end{figure}

Next, we consider the case with $K = 4$ users. We set $\alpha_k = 1/K, \forall k \in \{1,\ldots,K\}$, such that the $K$ users achieve a common average data rate. First, consider that the four users are uniformly distributed on the ground, with locations $(0,0)$, $(800/3~\textrm{m}, 0)$, $(1600/3~\textrm{m}, 0)$, and $(800~\textrm{m}, 0)$, respectively. Fig. \ref{f5} shows the optimized UAV trajectories under NOMA, FDMA, and TDMA, where $T = 100$ s and $V_{\textrm{max}} = 20$ m/s. It is observed that the optimized UAV trajectories under NOMA and FDMA each have two hovering locations that are above middle locations among the four users, while that under TDMA has four hovering locations each above one user. It is also observed that for TDMA, the UAV sequentially communicates with the nearest user, with the switching points dependent on the distances between the UAV and users. The observations are consistent with our findings in Remark \ref{r4}.

Fig. \ref{f6} shows the common average rate versus the communication duration $T$. It is observed that for all three schemes of NOMA, FDMA, and TDMA, as $T$ increases, the common average rates achieved by the proposed optimal solution and the successive-hovering-above-users scheme increase considerably, while those by the static-hovering scheme remain unchanged. It is also observed that for NOMA and FDMA, our proposed optimal solutions significantly outperform the benchmark schemes with successive hovering above users and static hovering; while for TDMA, the proposed optimal solution achieves the same performance as that by the successive-hovering-above-users scheme and outperforms the static-hovering scheme. It is further observed that the performance achieved by NOMA outperforms the performance by both FDMA and TDMA, while FDMA achieves higher performance than TDMA.

\begin{figure}[t]
\centering
\includegraphics[width=8cm]{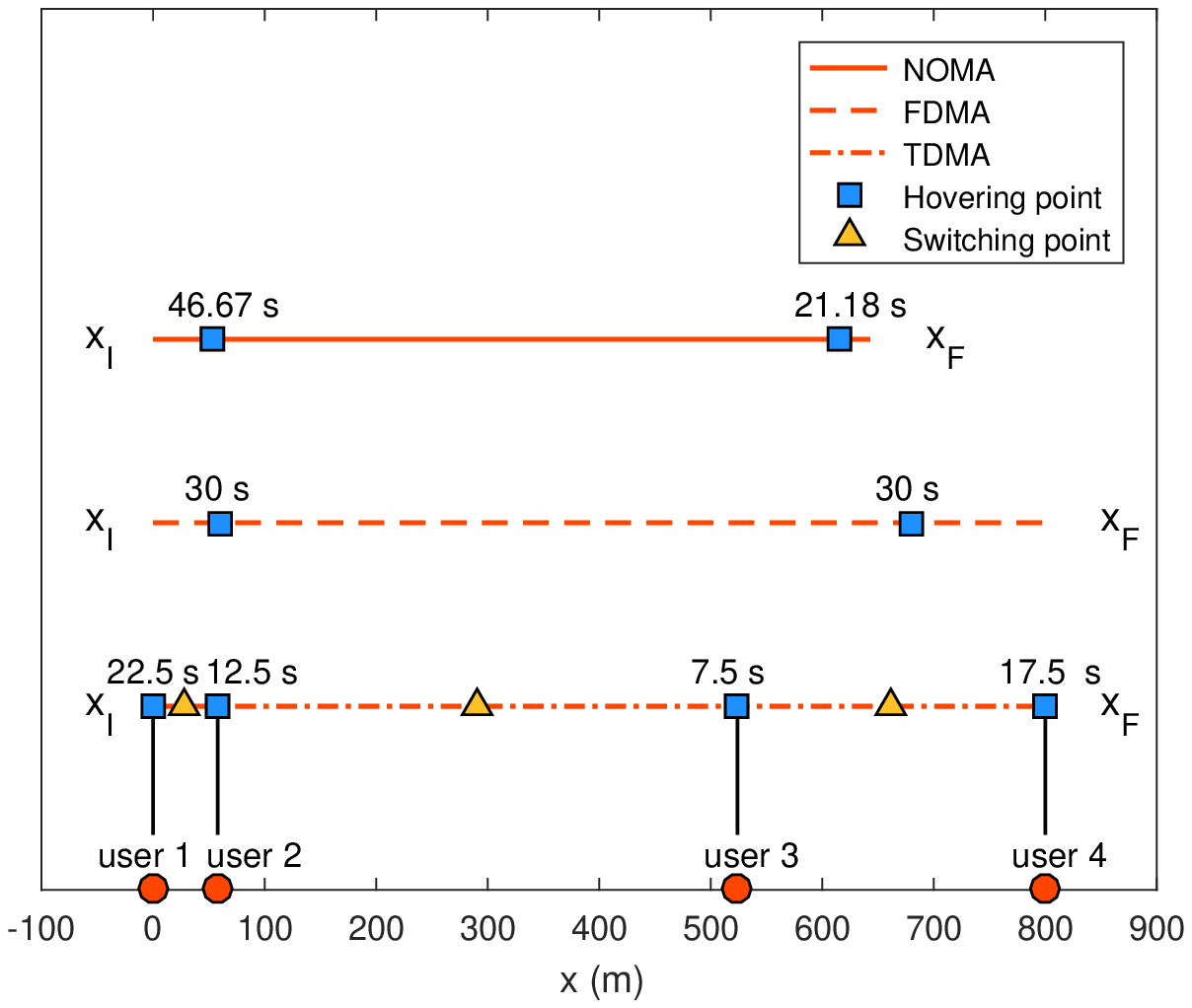}
\caption{Optimized UAV trajectories in the case with non-uniformly distributed users.}\label{f4}
\vspace{0.41cm}
\includegraphics[width=8cm]{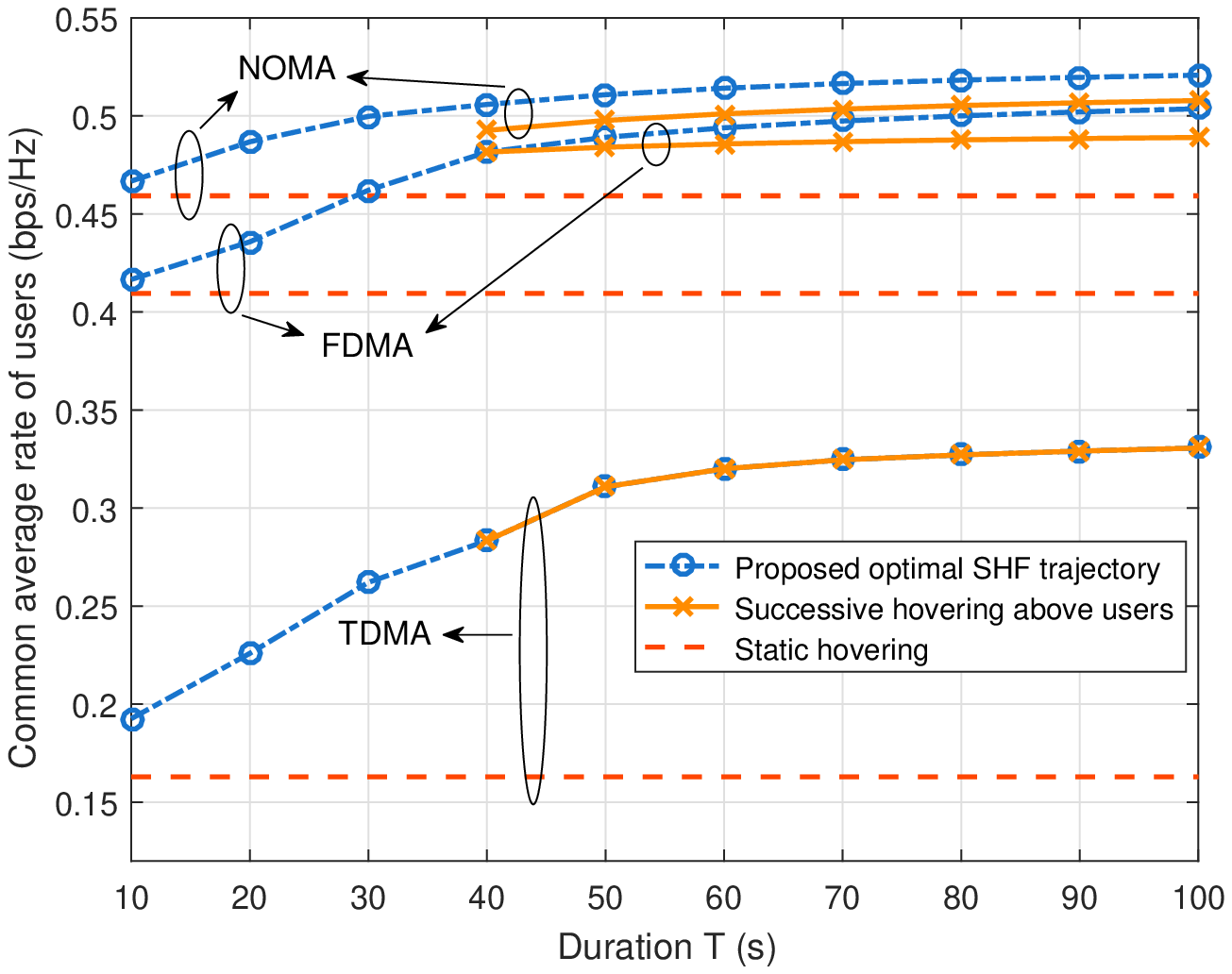}
\caption{Common average rate versus duration $T$ in the case with non-uniformly distributed users.}\label{f9}
\vspace{0.41cm}
\includegraphics[width=8cm]{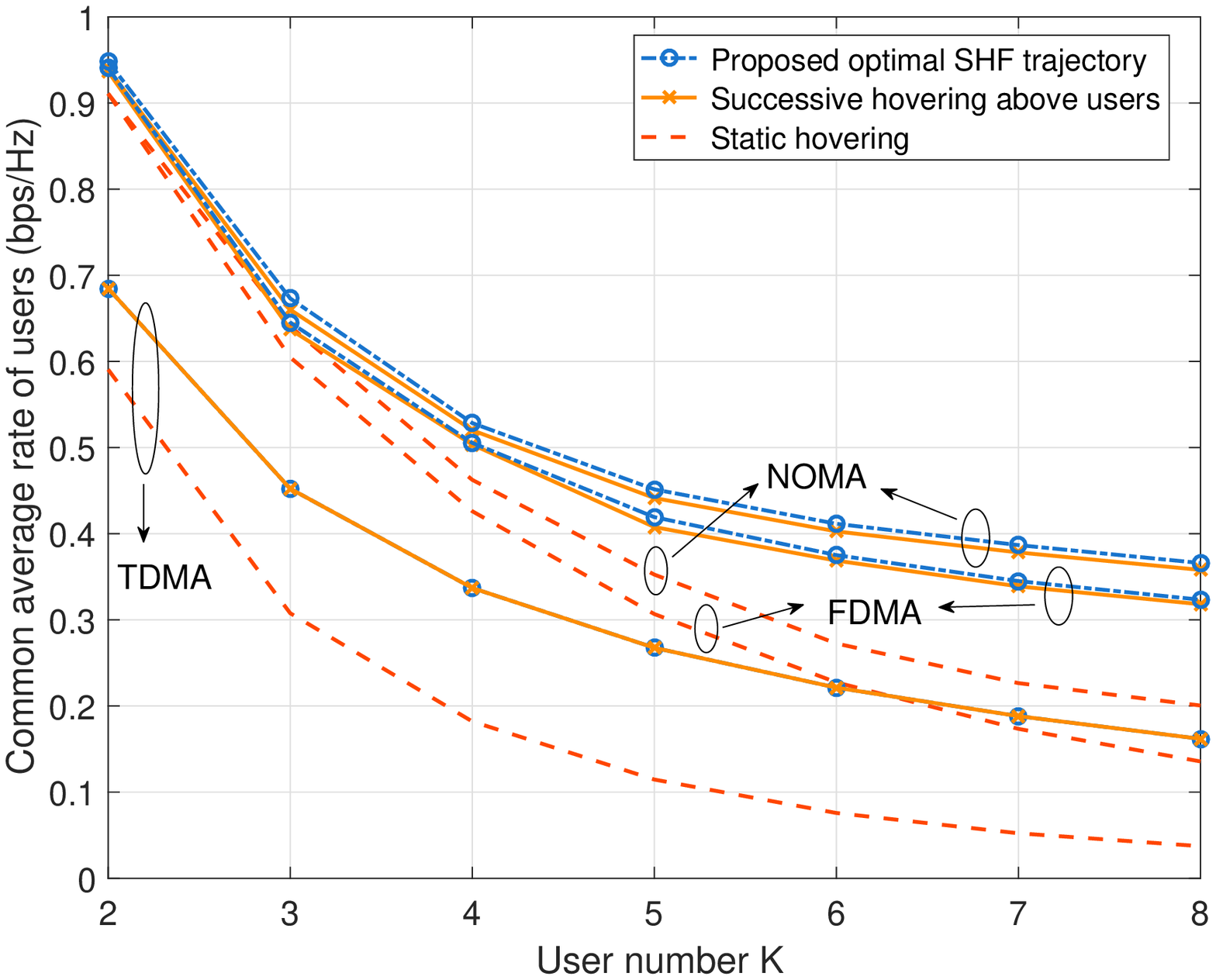}
\caption{Common average rate versus the number of ground users $K$ with $T=100$ s.}\label{f10}
\end{figure}

Then, we consider the case with four users non-uniformly distributed on the ground, at locations $(0,0)$, $(58~\textrm{m}, 0)$, $(524~\textrm{m}, 0)$, and $(800~\textrm{m}, 0)$, respectively. Fig. \ref{f4} shows the optimized UAV trajectories, under NOMA, FDMA, and TDMA, where $T\!=\!100$~s and $V_\textrm{max}\!=\!20$~m/s. Similarly as in Fig. \ref{f5}, it is observed that the optimal trajectories by NOMA and FDMA each have two hovering locations that are unsymmetrical, while that by TDMA has four hovering locations each above one user but with different hovering durations. Fig. \ref{f9} shows the users' common average rate. Similarly as in Fig. \ref{f6}, it is observed that for the cases with NOMA and FDMA, our proposed optimal solution significantly outperforms the two benchmark schemes; for TDMA, the proposed optimal solution achieves the same performance as that by the successive-hovering-above-users scheme and outperforms the static-hovering scheme. It is also observed that NOMA outperforms both FDMA and TDMA, while FDMA achieves much higher performance than TDMA.

Fig. \ref{f10} shows the common average rate versus the number of ground users $K$ with $T=100$ s, where the horizontal location of user $k\in {\mathcal K}$ is $w_k = ((200k-200)~ \textrm{m},0)$. It is observed that as $K$ increases, the achieved common average rates decrease, for all schemes under consideration. This is due to the fact that as $K$ becomes larger under our setup, the users are distributed in a larger area, and thus the far-apart users' channel conditions degrade. It is also observed that as $K$ increases, the proposed optimal-SHF-trajectory and successive-hovering-above-users schemes outperform the static-hovering scheme more significantly. This shows the significance of exploiting UAV mobility in enhancing the communication performance, especially in scenarios with a large number of distributed users.

\begin{figure}[t]
\centering
\includegraphics[width=8cm]{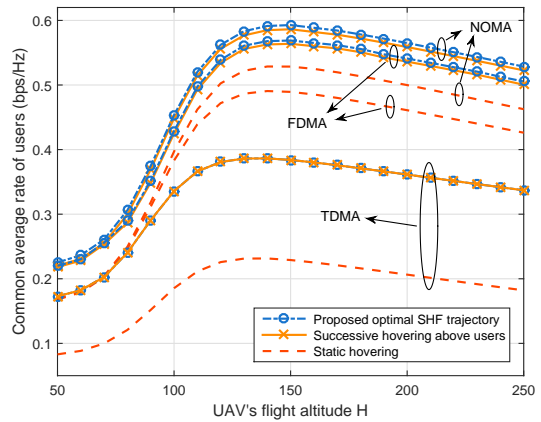}
\caption{Common average rate versus the UAV's flight altitude $H$ with $T=100$ s.}\label{f11}
\end{figure}

Fig. \ref{f11} shows the common average rate versus the UAV's flight altitude $H$ with $T=100$ s and $K=4$, where the four users' locations are shown in Fig. \ref{f5}. It is observed that for all schemes, the achieved common average rates first increase and then decrease, while the UAV's optimal flight altitude for maximizing the communication rate is generally different for each scheme. This is due to the fact that a higher altitude leads to higher LoS probability but longer distance between the UAV and each user, while a lower altitude corresponds to shorter distance but lower LoS probability, thus resulting in an interesting trade-off in designing the UAV's flight altitude for rate maximization. Similar phenomenon has also been observed in e.g. \cite{AA,you1}.
\section{Concluding Remarks}\label{sec:VI}
In this paper, we studied a UAV-enabled MAC, in which multiple users on the ground transmit individual messages to a UAV flying in the sky. By considering a linear topology scenario for ground users, we jointly optimized the 1D UAV trajectory and wireless resource allocation to reveal the fundamental rate limits of the UAV-enabled MAC over a particular communication period. In particular, we considered three transmission schemes, including capacity-achieving NOMA, as well as practical FDMA and TDMA. For each of the three schemes, we presented the globally optimal solution to the capacity/rate region characterization problem via convex optimization techniques, by showing that any speed-constrained UAV trajectory is equivalent to the combination of a maximum-speed flying trajectory and a speed-free trajectory. It was shown that the optimal 1D trajectory solutions follow an interesting SHF structure, but with different hovering locations and wireless resource allocation strategies under each multiple access scheme. Finally, numerical results showed that the proposed optimal trajectory design achieves considerable rate gains over other benchmark schemes, and the capacity region achieved by NOMA significantly outperforms the rate regions by FDMA and TDMA. Due to space limitation, there remain several interesting issues unaddressed in this paper, which are briefly discussed in the following to motivate future work.
\begin{figure*}
\begin{align} \label{62}
\int_0^{T}\!\log_2\!\left(1\!+\!\sum\limits_{k\in \bar{\mathcal K}}\frac{Ph_k({x}(t))}{\sigma^2}\right)\textrm{d}t=\!\int^{\bar{T}}_0\!\log_2\!\left(1\!+\!\sum\limits_{k\in \bar{\mathcal K}}\frac{Ph_k(\bar{x}(t))}{\sigma^2}\right)\!\textrm{d}t+\int^{\hat{T}}_0\log_2\!\left(1\!+\!\sum\limits_{k\in \bar{\mathcal K}}\frac{Ph_k(\hat{x}(t))}{\sigma^2}\right)\!\textrm{d}t,\ \forall\bar{\mathcal K}\subseteq{\cal K}.
\end{align}
\end{figure*}

{\it Extension to 2D and Three-dimensional (3D) Trajectory Design:} This paper considered the linear user topology with 1D UAV trajectory design for mathematical tractability and gaining essential design insights. In many practical applications, however, the users may be deployed irregularly in 2D or even 3D spaces. In this case, the corresponding 2D/3D UAV trajectory design problem will become very difficult to be solved optimally, due to the involvement of more optimization variables (i.e., the UAV’s 2D/3D locations over time), and more sophisticated speed constraints (e.g., the 2D horizontal trajectory constraints together with the vertical ascending and descending speed constraints). More specifically, although we may be able to prove that the optimal 2D/3D trajectory still follows the SHF structure (based on a similar proof technique as that in Section \ref{sec:III}), it becomes very difficult to find the optimal maximum-speed flying trajectory as it is non-straight in the 2D/3D space. As such, how to obtain the globally optimal 2D/3D UAV trajectory solution to the capacity region characterization problem is an open problem that is worth pursuing for future work.

{\it Energy-Efficient UAV Communications with Capacity-Achieving NOMA:} There generally exists a fundamental trade-off between communication rate and energy consumption in UAV communications. In practice, the UAV energy consumption consists of two parts for communication and propulsion, respectively \cite{sic1,zeng4}. How to design energy efficient UAV communications with capacity-achieving NOMA for multiple on-ground communicating users is also an interesting and unaddressed problem.

{\it UAV-enabled BC:} Although this paper focused on the UAV-enabled MAC, the design principles are extendable to the UAV-enabled BC, based on the well-established MAC-BC duality \cite{dual}. Different from the UAV-enabled MAC with individual power constraints at on-ground user transmitters, the UAV-enabled BC is subject to the sum power constraint at the UAV transmitter. How to jointly optimize the UAV trajectory design and transmit power allocation is a new problem worth further investigation.

\appendices
\section{Proof of Lemma \ref{l2}} \label{AppA}
To start with, we divide the total period $\mathcal T$ into $N$ sub-periods each with identical duration $\delta = T/N$, where each sub-period $n \in {\cal N} \triangleq \{1,\ldots, N\}$ is denoted as $\mathcal{T}_n = ((n-1)\delta,n\delta]$. Accordingly, the trajectory ${x(t)}$ is partitioned into $N$ sub-trajectories. Here, $N$ is chosen to be sufficiently large such that the UAV flies at an approximately constant speed $v_n \le V_{\textrm{max}}$ over each sub-period $n$, i.e., $x(t) = x((n-1)\delta) + v_n(t - (n-1)\delta), \forall t\in{\mathcal T}_n$. In the following, we construct the maximum-speed flying trajectory $\{\bar x(t)\}$ and the speed-free trajectory $\{\hat x(t)\}$ by considering each of the $N$ sub-periods, respectively.

Consider one particular sub-period ${\cal T}_n, n\in{\cal N}$. First, we construct the maximum-speed sub-trajectory, in which the UAV flies from the initial location $x((n-1)\delta)$ to the final location $x(n\delta)$ over this sub-period at the maximum speed $V_\textrm{max}$, with the required duration being $\bar{\delta}_n ={\delta v_n}/{V_{\textrm{max}}}$. Accordingly, we define the sub-period as ${\cal \bar{T}}_n=\left(\sum_{i=1}^{n-1}\bar{\delta}_i,\sum_{i=1}^{n}\bar{\delta}_i\right]$, and the corresponding sub-trajectory as $\bar{x}(t)=x((n-1)\delta)+V_{\textrm{max}}(t-\sum_{i=1}^{n-1}\bar{\delta}_i), \forall t\in\bar{\cal T}_n$. Next, we construct the other sub-trajectory, in which the UAV flies from the initial location $x((n-1)\delta)$ to the final location $x(n\delta)$ of this sub-period at speed $\hat{v}_n={\delta v_n}/{\hat{\delta}}$, where $\hat{\delta}_n=\delta-\bar{\delta}_n$ denotes the required duration. Accordingly, we define the sub-period as ${\cal \hat{T}}_n=\left(\sum_{i=1}^{n-1}\hat{\delta}_i,\sum_{i=1}^{n}\hat{\delta}_i\right]$, and the corresponding sub-trajectory is $\hat{x}(t)=x((n-1)\delta)+\hat{v}_n(t-\sum_{i=1}^{n-1}\hat{\delta}_i),\forall t\in\hat{\cal T}_n$. Notice that in the special case with $v_n = 0$, we have $\bar{\delta}_n=0$ for the maximum-speed sub-trajectory; while in the other special case with $v_n = V_{\textrm{max}}$, we have $\hat{\delta}_n=0$.

Now, by combining the sub-trajectories $\{\bar{x}(t)\}$ and $\{\hat{x}(t)\}$ over the $N$ sub-periods $\bar{\cal T}_n$'s and $\hat{\cal T}_n$'s, we obtain two trajectories $\{\bar{x}(t)\}$ and $\{\hat{x}(t)\}$ with total durations $\bar T = (x_{\textrm{F}} - x_{\textrm{I}})/T$ and ${\hat T} = T - {\bar T}$, respectively. It is clear that $\{\bar{x}(t)\}$ corresponds to a maximum-speed flying trajectory from $x_{\textrm{I}}$ to $x_{\textrm{F}}$, while $\{\hat{x}(t)\}$ generally does not satisfy any speed constraints and thus is named as a speed-free trajectory. Furthermore,
it is evident that by combining the two constructed trajectories $\{\bar{x}(t)\}$ and $\{\hat{x}(t)\}$ together, the UAV visits the same locations (with the same duration at each location) as in the original trajectory $\{x(t)\}$. Therefore, it corresponds to \eqref{62} at the top of the this page. By comparing $\bar{\mathcal C}(\{x(t)\})$ in \eqref{4} and $\hat{\mathcal C}(\{\bar x(t)\},\!\{\hat{x}(t)\})$ in \eqref{111}, we have $\bar{{\mathcal C}}(\{x(t)\})=\hat{{\mathcal C}}\left(\{\bar x(t)\},\{\hat x(t)\}\right).$

As a result, Lemma \ref{l2} is finally proved.

\end{document}